# "SECULAR LIGHT CURVE OF EXOCOMET 3I/ATLAS, AND ITS LOCATION ON A COMET EVOLUTIONARY DIAGRAM"

**Ignacio Ferrín[1], José Garrido[2], Charles Triana[3], Giuliat Navas[4], Raul Melia[5], Santiago Pérez[6], Emiliano Gómez[6], Jorge Andrey Vargas[6], Juan Hincapie[7], Brayan Quintero[8]**

**1- University of Antioquia, Institute of Physics, FACOM-SEAP, Medellin Colombia.**

**ignacio.ferrin@udea.edu.co**

**2- International University of Valencia, VIU, Spain**

**3- AstroExplor Observatory, W60, Boyacá, Colombia**

**4- Astronomy Research Center CIDATA, Mérida, Venezuela**

**5- Raul Melia Observatory, Pekos Planetary, Carl Sagan Planetary, Carlos Paz, Argentina.**

**6- University of Antioquia, Astronomy Program, Medellin Colombia.**

**7- University of Antioquia, Physics Program, Medellin Colombia.**

**8- Instituto de Educación Técnica Profesional, Roldanillo, Colombia.**

**Number of Pages: 41**

**Number of Tables: 3**

**Number of Figures: 23**

**Number of Citations: 65**

## Highlights

1- The colors of the 3I exocomet are consistent with the colors of solar system comets.

2- There is evidence to believe that the object is a binary.

3- This exocomet is 93% $CO_2$, dry ice.

4- The object has a Mass-Loss-Age of 0.16 comet years. In a Comet Evolutionary Diagram it is a baby comet.

5- This exocomet belongs to the Oort Cloud but from another stellar system.

## Abstract

In this work we will create the Secular Light Curve (SLC) of exocomet 3I/ATLAS, using the SLC-Methodology (Ferrín 2010-2023). The SLCs give a throve of new information and allow the comparison of exo-comets with comets of our own solar system. We arrive at the following conclusions: The colors of 3I are consistent and lie inside the area of colors of other comets in our solar system. The SLC of this comet exhibits a photometric anomaly, a region from -120 to -45 days before perihelion that we interpreted as an eclipse, suggesting that 3I might be a binary. At -45 days, the SLC changes abruptly its slope, reaching a maximum absolute magnitude of $m_V(1,1,\alpha) = 6.8 \pm 0.1$. Using reported estimates derived from the scrutiny of 97 papers in the arXiv.org depository for the size, dust, $H_2O$, $CO_2$, and CO production rates, we calculate the total mass loss. We use the inverse total mass loss, as a proxy for age, since comets lose mass as they age. The Mass-Loss Age = 0.16 comet years (a baby comet) will be plotted in the horizontal axis of a Comet Evolutionary Diagram (CED) while the number of Remaining Returns defined as $RR = r/\Delta r = 24$, will be plotted in the vertical axis of the CED. 3I/ATLAS exocomet lies among the comets of our Oort cloud family. We conclude that 3I is a comet of the Oort Cloud, but from a different stellar system. The Evolutionary Diagram presented in this work shows complexity beyond current understanding.

# 1- INTRODUCTION

Comet 3I/ATLAS has emerged as a singular object of astronomical study, being the third exo-object identified after 1I/Oumaumau and 2I/Borisov (Guzik et al., 2020; Jewitt et Luu, 2020). 3I/ATLAS was discovered on July 1, 2025, by the Asteroid Terrestrial-impact Last Alert System (ATLAS) telescope in Río Hurtado, Chile (Denneau, L. et al., 2025). Its discovery offered a unique opportunity to investigate the composition and evolution of minor bodies originating outside our Solar System, providing a crucial context for understanding planetary formation across the Milky Way (Trilling et al., 2017). Exocomets are time capsules that retain some pristine information from their system of origin, making the detailed analysis of 3I/ATLAS's light curve fundamental to comparative astrophysics. In particular, the Secular Light Curves (SLC) Methodology (Ferrín 2005a, 2005b, 2006, 2007, 2008, 2009, 2010a, 2010b, 2012, 2013, 2014a, 2014b, 2017, 2018, 2019), have proven to be a powerful tool for revealing the internal physical characteristics and outgassing processes of comets. This allows for a direct comparison of their evolutionary behavior with that of comets from our own cosmic neighborhood (Mumma and Charnley, 2011). This work focuses on applying the SLC analysis to 3I/ATLAS to determine its evolutionary parameters, assess its stability, and explore the surprising hypothesis, based on the Comet Evolutionary Diagram (CED), that this object might come from the Oort Cloud but coming from a different stellar system.

Up to 2025 eight comets have been visited by spacecraft, 1P/Halley, 21P/Giacobini-Zinner, 26P/Grigg-Skjellerup, 19P/Borrelly, 81P/Wild 2, 9P/Tempel 1, 103P/Hartley 2, and 67P/Churyumov-Gerasimenko. The amount of scientific information that has been gathered is astounding. Additionally, the two space telescopes, Hubble and James Web, are generating data previously unreachable with ground-based telescopes. And the Vera Rubin telescope has just discovered 11000 asteroids. We have learned that the nucleus of comets looks like potatoes, that their surfaces are very black (albedo 0.02 to 0.07), that their D/H ratio does not agree with that of Earth, and that there are complex molecules on their surfaces, some of which are considered as life precursors.

Despite all this progress, we are still missing something of great importance, and that is a Comet Evolutionary Diagram (CED). Stellar astrophysics has a very famous plot, the Hertzsprung-Russell diagram (Russell, 1914), that shows in the vertical axis the stellar luminosity, and in the horizontal axis the temperature of the star. The importance of this diagram is that stars move on the plot, depending on their physical parameters, and from models their age can be calculated.

In 2005 the first author asked a question: if stars had an evolutionary diagram, why comets didn't? (Ferrín, 2012). It took the author several years to determine what the horizontal and vertical axis could look like, since comets do not emit light and they do not have nuclear reactions in their cores (Ferrín et al, 2012, 2013, 2014, 2019). In this work we will place comet 3I/ATLAS in a Comet Evolutionary Diagram, CED, to learn its relationship with our own comets.

## 2- Methodology and data reduction

We employed the Secular Light Curve Methodology (Ferrín, 2005a to 2019) to study the cometary activity of 3I/ATLAS. This methodology specifies that the light curve must be done fulfilling some specific rules that we will use in this work. We extracted photometric observation from the Minor Planet Center (MPC) APIs, which provides data such as date, the distance Earth-object, Δ (the geocentric distance), the distance object-Sun R (heliocentric distance), phase angle, α, observed magnitude and filter. We also used data from COBS that contains visual observations of great value (Comet Observations Database, https://www.cobs.si/ ).

In that formalism we adopt the envelope of the dataset as the correct interpretation of the light curve. There are many physical factors that affect comet observations, like twilight, moonlight, haze, cirrus clouds, dirty optics, lack of dark adaptation, excess magnification, and in the case of CCDs, sky background too bright, insufficient time exposure, insufficient CCD aperture, and a too large scale. All these factors decrease the captured photons coming from the object, and the observer makes an error downward, toward fainter magnitudes. There are no corresponding physical effects that could increase the perceived brightness of the object. Thus, the envelope is the correct interpretation of the light curve. In fact, the envelope is flat, while the anti-envelope (the fainter magnitudes of the distribution), is diffused and irregular. The envelope represents an ideal observer (vision 20/20), using an ideal telescope and detector, in an ideal atmosphere (pure and transparent).

The MPC Observations Database is an astrometric database that uses small photometric apertures to extract the flux. Thus, the MPC will produce fainter magnitudes than COBS because COBS contains many more visual observations capable of extracting the whole coma flux.

The problem of a small observing aperture will resurface when we deal with the production rates of the subliming components, H2O, CO2, CO and dust. These production rates will show large scatter and, in many cases, will not be able to extract the whole flux from the coma. According to Harrington Pinto (2022), CO2, CO and H2O can rarely be detected simultaneously using the same instrumentation. So, the measurements are derived using a mixing of different techniques and calibrations, making it difficult to derive these quantities with certainty.

To make the envelope sharp, we reduce all the filtered observations to the V-band using the transformation equations of Jordi et al. (2006). In this way we can reduce the uncertainty of the envelope to ~±0.15 magnitudes, our detection limit. In Figure 1 we show the flatness of the light curve after the filter correction.

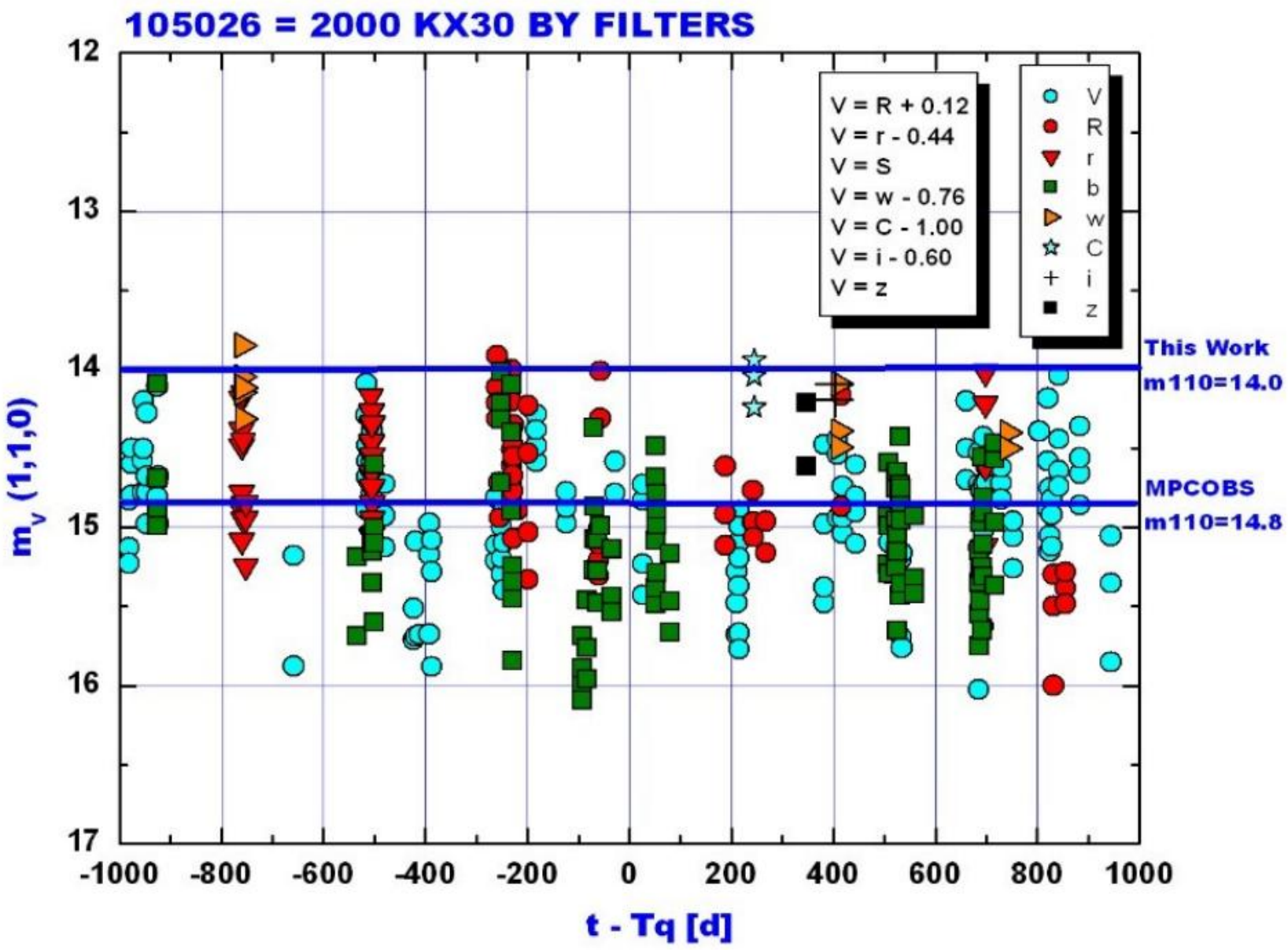

Figure 1. Effect of reducing the filtered observations of asteroid 2000 KX30 to the V-band. The scatter of the data points has been reduced to about ~±0.15 magnitudes. The MPC absolute magnitudes are fainter than observed by 0.8 magnitudes. There is a systematic error in all the MPC absolute magnitudes, because they do not use the envelope of the observations.

The reduced absolute magnitude in the visual band, $m_V(1,1,0)$, is calculated from the observed magnitude $m_V(\Delta,R,\alpha)$ for each measurement, correcting for geometric effects using the equation

$$m_V\,(1,1,0) = m_V\,(\Delta,R,\alpha) - 5 \log\,(\Delta.R) - \beta\alpha \qquad (1)$$

where $\beta$ is the phase coefficient. All the photometry was reduced to the V-band by means of color corrections taken from the MPC. If the object exhibits a cometary behavior with a coma, the nucleus cannot be seen and $\beta$ is undefined and left as $\alpha$.

From these values, phase diagrams and SLCs are constructed with respect to the time to perihelion, t - Tq, allowing the identification of potential brightness anomalies associated with cometary activity. This Methodology was implemented in a web tool developed in Python and Streamlit, which automates data download, processing, graphing, and report generation, resulting in a reproducible workflow suitable for optimal and careful deep analysis of each object.

The Secular Light Curve Methodology has been used successfully by several authors.

Nazarov (2024) studied comet 7P/Pons-Winnecke and found a turn-on Ron=1.76±0.1 au, and a photometric age Page=54.4 comet years, a moderately old comet. The light curve amplitude was Asec(1,1)=5.5 magnitudes, corresponding to the state of a transitional middle-age comet.

Rondon et al. (2025) studied the SLCs of 52 periodic comets observed by SWAN-SOHO. They found that 81P and 96P are the youngest and the oldest, and 55P and 96P are the comets with the greatest and smallest active regions.

Liang et al. (2023), used the Secular Light curves to study comets 133P/Elst-Pizarro, 176P/Linear, 288P/300163 and 433P/248370). They found that 176P had an amplitude of 0.37 magnitudes, but in 2011, 2017 and 2022 it did not show any activity. They also detected a change in the Ton and Toff values between 1996-2001 and 2007-2013, which basically means evolution. One of the most interesting conclusions of these authors is that they found evolution as a function of time. For 133P/Elst-Pizarro the amplitude of SLC was 2.4 mag in 1996-2001 perihelion passages but changed to 1.2 mag in 2007-2013 perihelion passages. They also found that 288P/(300163) had a large asymmetry parameter, Roff/Ron, which means that the sublimation takes place in depth, not just in the surface.

Liang et al. (2025) collected the observations and created the SLCs of comets 238P/Read, 259P/Garrad, 313/Gibbs, 324/La Sagra and 358P/PANSTARRS, and determined their age though the parameter Page.

Li et al. (2023) analyzed the SLCs of comets 60P/Tsuchinshan 2 and 62P/Tsuchinshan 1. They found that both comets could be classified as middle-age, and that their perihelion distances are decreasing, placing them into the class of Lazarus comets who are rejuvenating.

The above information and much more was generated using the SLC Methodology.

## 3- Telescopes

Observations of interstellar comet 3I/ATLAS were conducted using three distinct facilities: The Astro Explor site (W60) and the Carlos Paz Observatory, both utilizing 25 cm telescopes operated by C. Triana and R. Melia, respectively, and the National Astronomical Observatory of Venezuela (MPC code 303), where G. Navas employed the 1.0 m Schmidt telescope. All captured data underwent standard reduction procedures, including bias, dark, and flat-field corrections, to ensure photometric and astrometric precision.

# 4- Color-Color Diagrams: Is it really a comet?

Using the photometric data published in the literature by Bolin et al. (2025), Alvarez-Candal (2025), and Opitom et al. (2025) (Table 1 and Figures 2 and 3) it is possible to create the color-color diagrams, in which comets occupy a phase space.

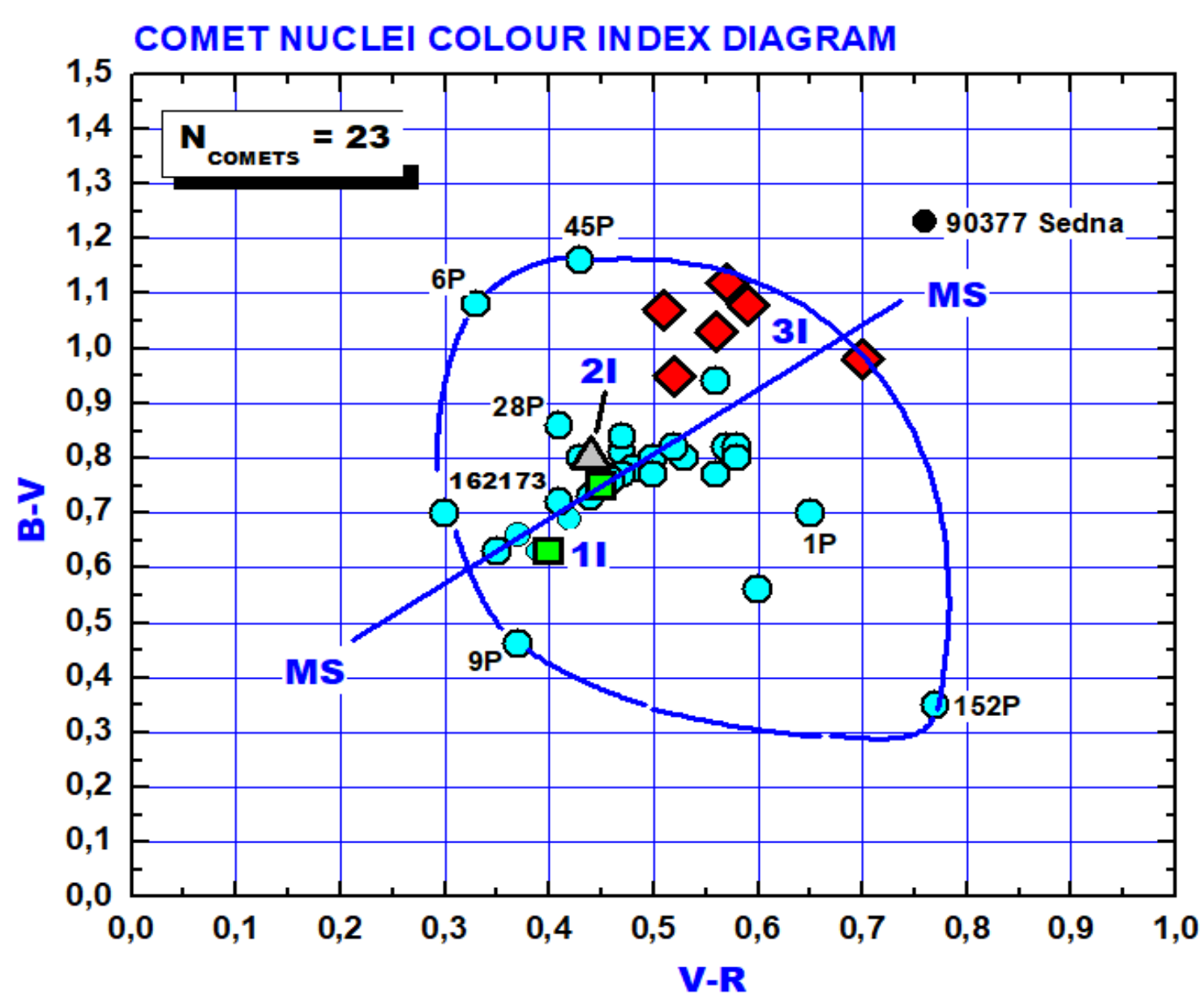

Figure 2. Color-color diagram (B-V vs V-R) for 23 comet nuclei, illustrating the photometric diversity of cometary surfaces compared to solar system and interstellar objects. 70% of the objects lie on the solid diagonal line (Main Sequence, MS) indicating the solar color locus. The distribution highlights the positions of interstellar visitors, including 1I/Oumuamua (green square), 2I/Borisov (grey triangle), and 3I/ATLAS (red diamond cluster), showing their varying degrees of "redness" relative to the typical cometary population enclosed within the blue elliptical boundary. Short-period comets such as 1P/Halley, 9P/Tempel 1, 6P/d'Arrest, and 45P/Honda-Mrkos-Pajdušáková are identified to demonstrate surface composition variety, while the trans-Neptunian object 90377 Sedna (upper right) serves as a benchmark for a non-cometary extremely red surface.

Although the colors for 3I/ATLAS show some dispersion, they remain strictly within the established cometary domain, confirming its classification as a comet. In contrast, the trans-Neptunian object (90377) Sedna is plotted as a reference to illustrate a non-cometary photometric profile, of extremely red surface.

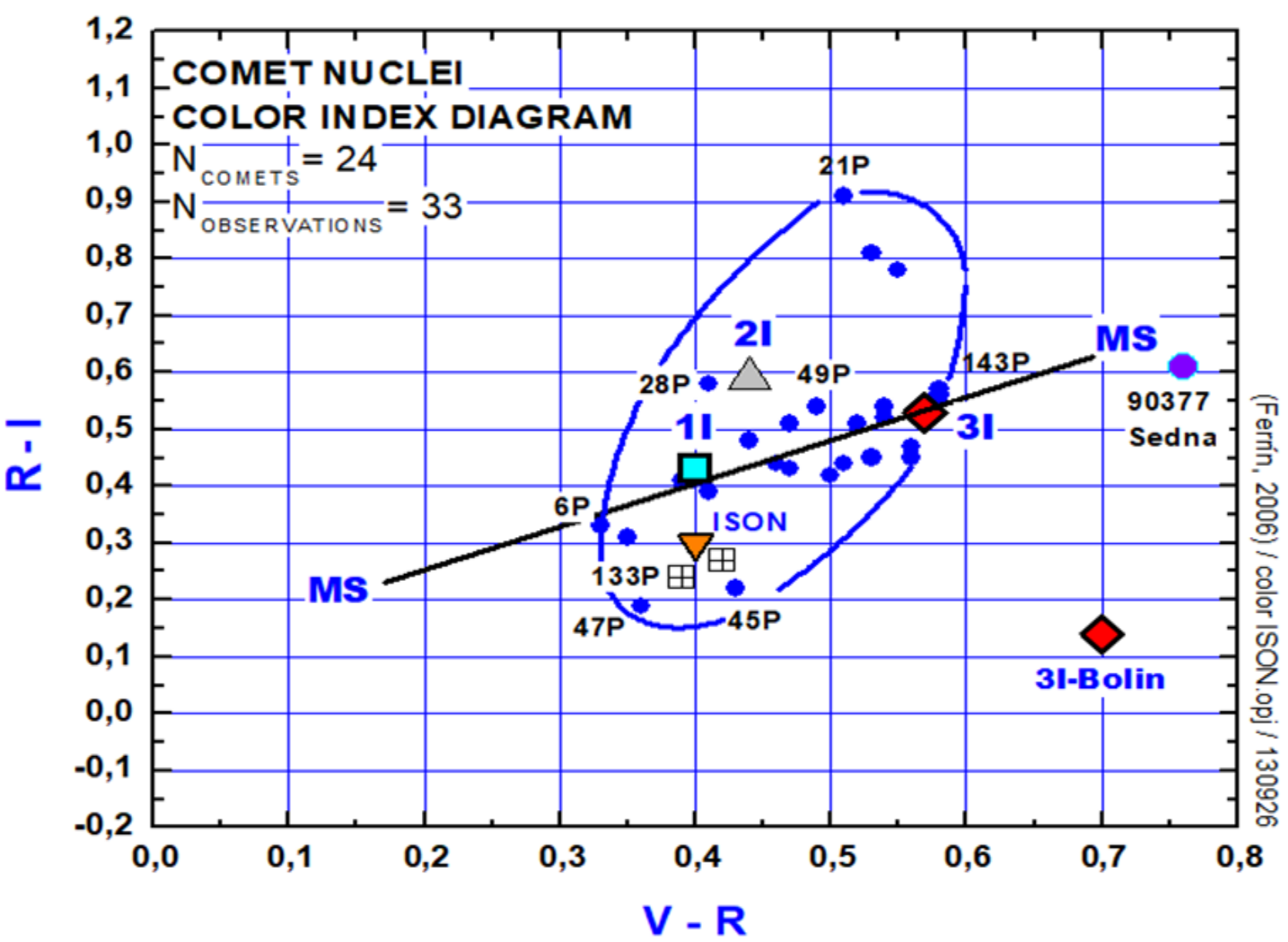

Figure 3. Color-color diagram (R-I vs V-R) for 24 comet nuclei based on 33 photometric observations, where the black diagonal line denotes the Main Sequence (MS) or solar color locus. The distribution highlights the positions of interstellar objects 1I/Oumuamua (cyan square), 2I/Borisov (grey triangle), and 3I/ATLAS (red diamonds), with the 3I-Bolin data point exhibiting a notable departure from the standard cometary population enclosed within the blue ellipse. Notice that 1I and 3I lie on the Main Sequence. For comparative context, several periodic comets are identified (21P, 28P, 45P, 143P), alongside the ultra-red TNO 90377 Sedna (upper right) and the comet C/2012 S1 (ISON), indicated by an orange triangle.

There have been some works in the literature suggesting the 1I was not a comet. The two diagrams B-V vs V-R and R-I vs V-R dispel that idea.

The plotted values for 1I, 2I and 3I fall within the expected region for primitive bodies (Figures 1 and 2), confirming their classification as cometary objects. The proximity of 3I/ATLAS to the MS suggests a surface dominated by the same primitive, organic-rich materials found in the Solar System's own small bodies.

Table 1. Measured colors of exocomet 3I/Atlas.

| Author | B-V | V-R | R-I |
|---|---|---|---|
| Bolin et al. (2025) | 0.98±0.23 | 0.71±0.09 | - - - - - - - - |
| Alvarez-Candal (2025) | 1.07±0.10 | 0.51±0.06 | - - - - - - - - |
| Opitom et al. (2025) | 1.12±0.14 | 0.57±0.09 | 0.53±0.17 |

The color indices reported by Bolin et al. (2025), Alvarez-Candal (2025), and Opitom et al. (2025), show strong internal consistency (Table 1). The average values of B-V ≈ 1.05 and V-R ≈ 0.60 indicate that 3I/ATLAS exhibits a significantly redder spectrum than the Sun, a characteristic typical of cometary nuclei and primitive Solar System objects rich in organic material. While minor variations exist in the R-I index across the different studies, the data confirms that the exocomet's spectral slope does not present any extreme chromatic anomalies.

In conclusion, the photometric properties of 3I/ATLAS are entirely compatible with those of short- and long-period comets in our Solar System. This suggests that, despite its interstellar origin, the surface materials and chemical evolution processes in its system of origin are analogous to those observed in the Oort Cloud and the Kuiper Belt.

## 5- Larson-Sekanina Filter: Is it falling apart?

Larson and Sekanina (1984) developed a very powerful algorithm to increase the resolution of comet images. We are going to use it to study the inner region of 3I searching for jets. The results are surprising and appear in Figures 4, 5 and 6.

Images taken with a SW 250 mm reflect or telescope coupled with a ZWO QHY174M camera, R. Melia took 10 images of the object on November 17, 2025, 18 days after perihelion, when the comet was at 10° altitude, the seeing was 2.4" of arc FWHM and the distance to the Sun R= 2.056 AU.

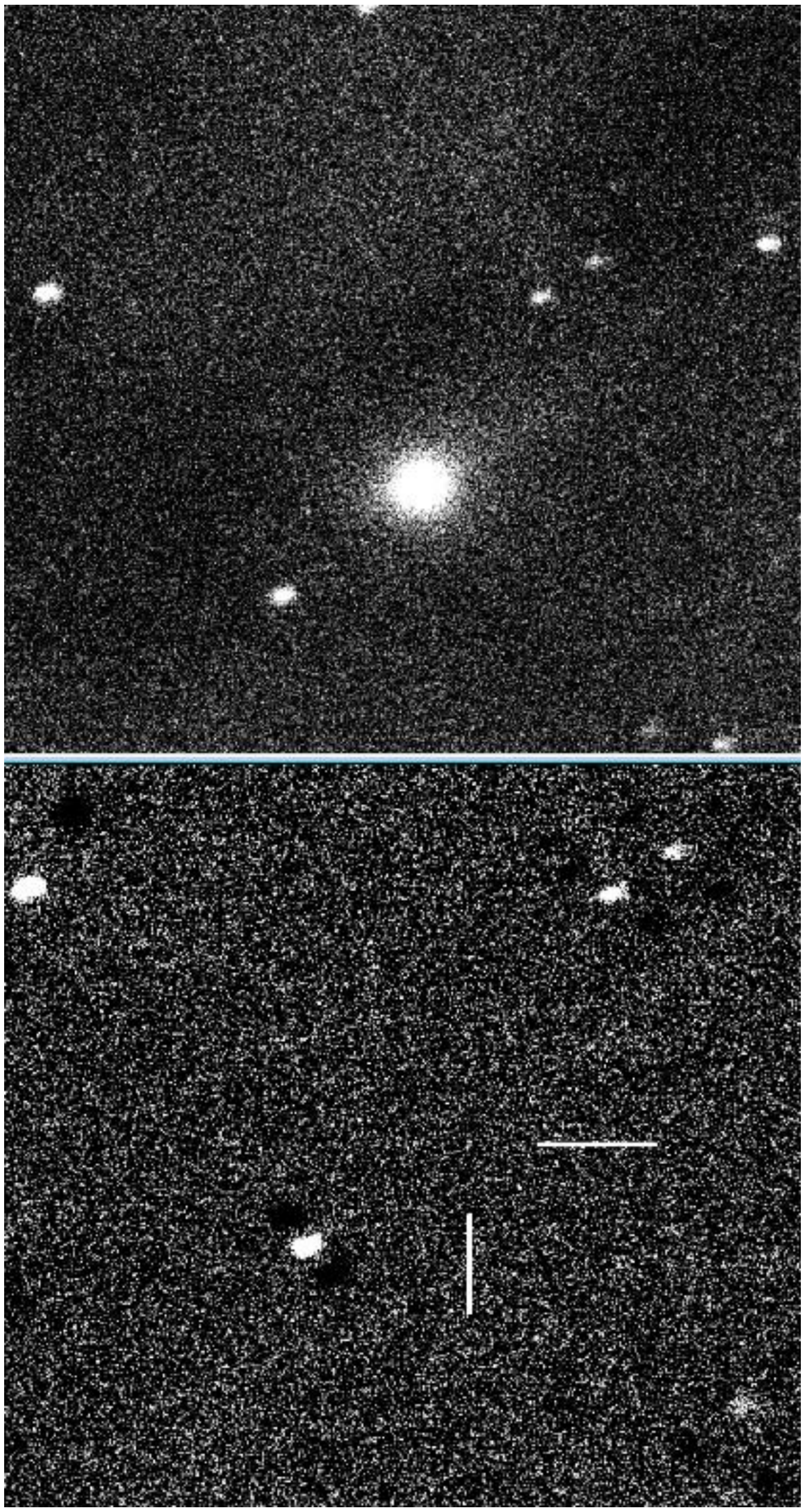

Figure 4. No trail of debris no comet. In the upper panel we show the original image of the comet taken on November 17, 2025, 18 days after perihelion, when R=1.516 au. On the lower panel, we apply the Larson-Sekanina filter with a 10° rotation. The comet disappears completely, a most estrange behavior for a normal comet. This result implies that there is not any significant jet that could be the result of sublimation. Additionally, the coma is a perfectly symmetric 3-dimentional gaussian implying that sublimation takes place all over the area homogeneously, a perfect symmetrical cloud. The rotational symmetry of the coma was also mentioned by Lisse et al. (2026a).

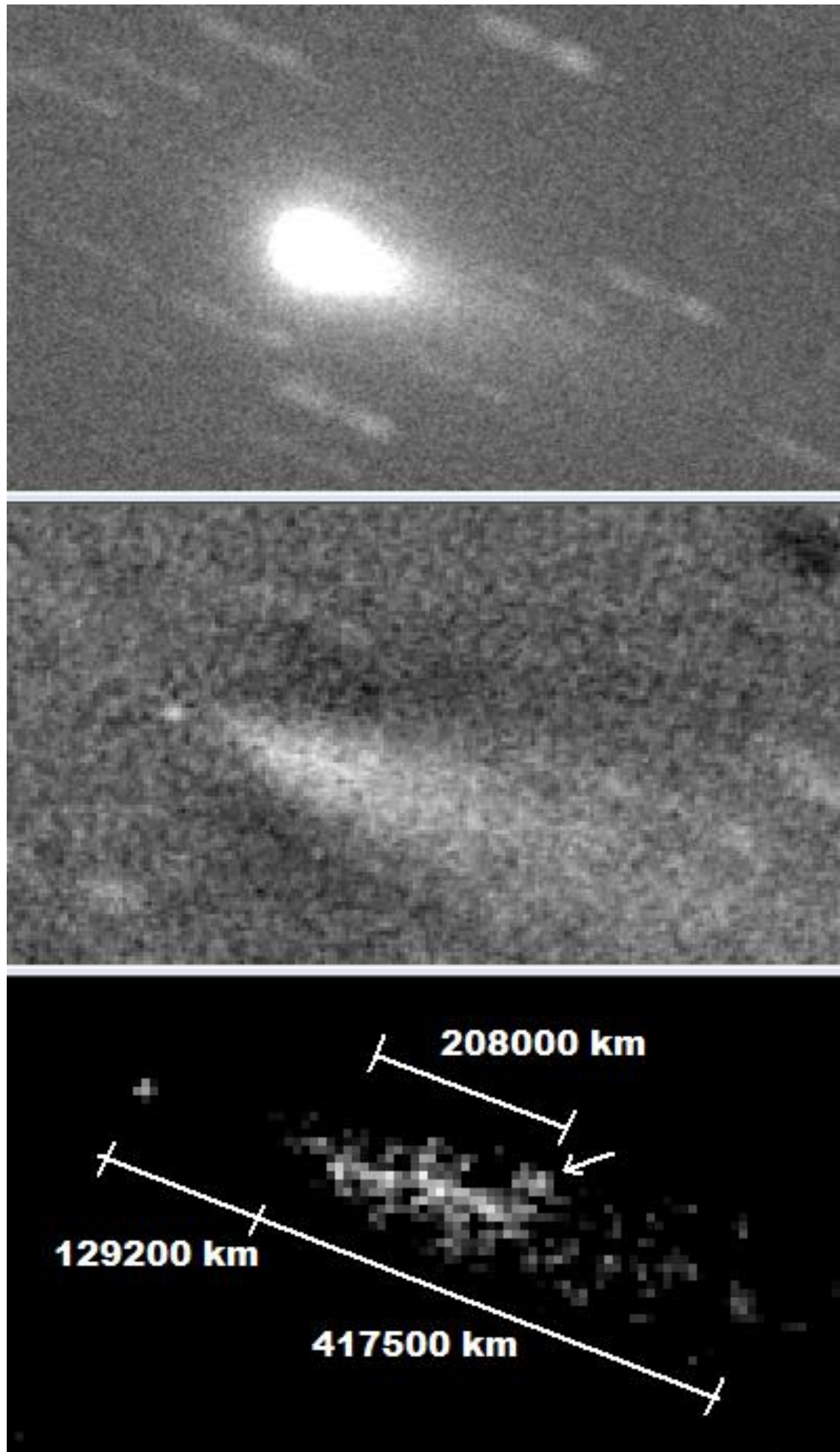

Figure 5. Trail of debris. Larson-Sekanina filter applied to an image of 3I taken on 2025 12 31, by C. Triana, with a 25 cm telescope 61 days after perihelion when R=2.637 au. 79 images of 30 s each for a total exposure time of 39.5 minutes were stacked using average filter. Upper image is the comet. Middle image is the L-S filter with a rotation of 20°. Lower image is full stretching of the middle image. Notice the long filament in the tail. Distances are projected in the sky. In the lower image notice the boulder in the upper right hand of the tail. Also notice the blackness of the surrounding sky, meaning that all these pixels are above the sky background. Each pixel measures ~7500 km. 414500 km is 1.09 the distance to the moon.

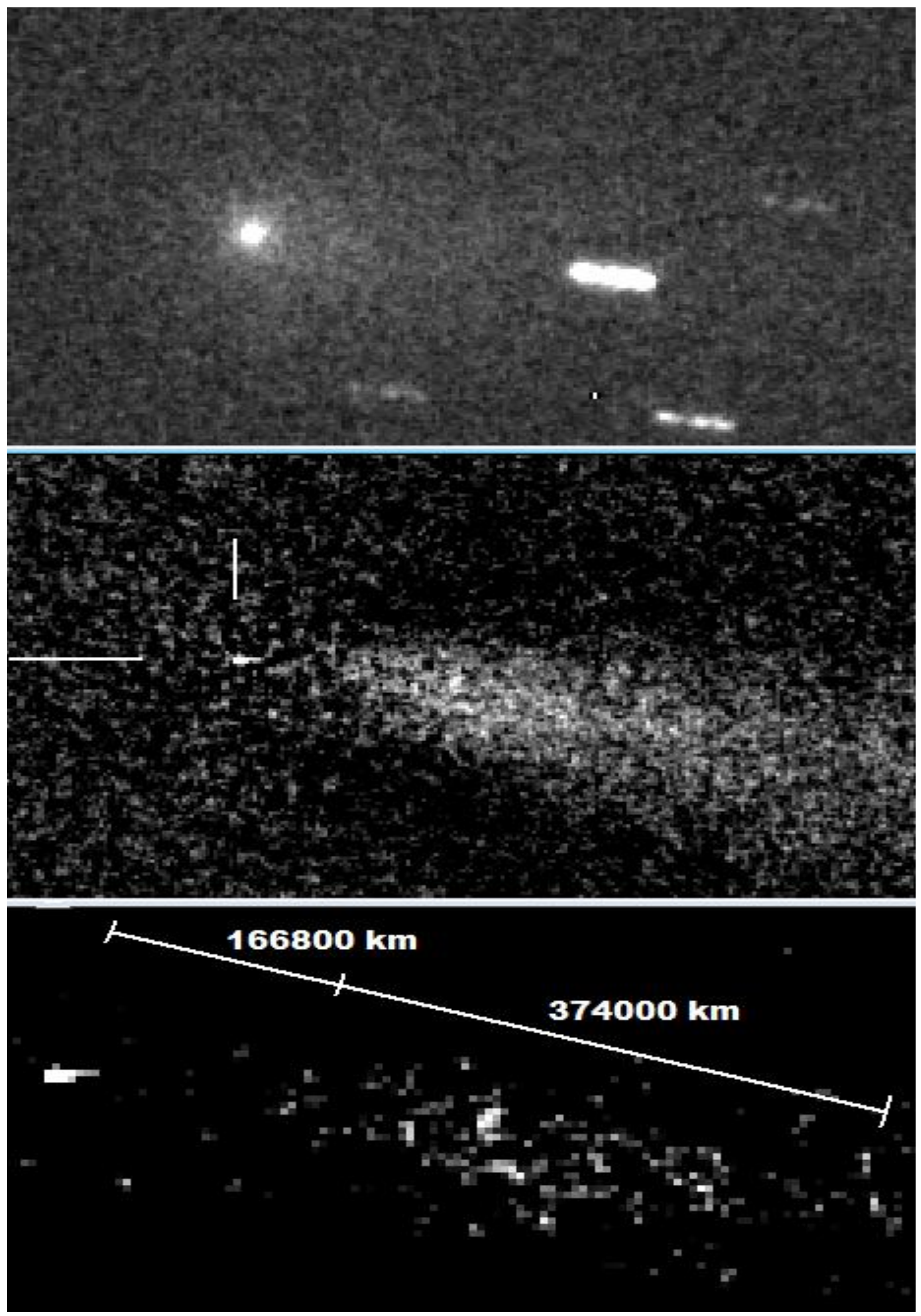

Figure 6. Trail of debris. Larson-Sekanina filter applied to an image of 3I taken on 2026 01 22, by G. Navas, with a 1 m Schmidt telescope of the National Observatory of Venezuela, 83 days after perihelion when R=3.33 au. 12 images with total exposure time of 2.5 minutes were stacked using the average filter. Upper image is the comet. Middle image is the L-S filter with a rotation of 20°. Lower image is full stretching of the middle image. Notice a jet 6565 km long (half the Earth radii), coming out of the nucleus, and a large boulder in the tail. Distances are projected in the sky. Each pixel measures ~5200 km. 374000 km is almost the distance to the moon.

The application of the Larson-Sekanina filter to observations of 3I/ATLAS reveals a highly unconventional morphological evolution, characterized by a lack of traditional jet activity and a transition toward macroscopic mass shedding. Initial analysis from November 2025 shows a perfectly symmetric, Gaussian coma devoid of sublimation-driven features, as evidenced by the object's complete disappearance under rotational filtering—a result consistent with the findings of Lisse et al. (2025a). However, subsequent filtered images from late 2025 and early 2026 highlight the emergence of a 6,565 km jet and distinct, boulder-sized fragments within a projected trail of debris. These results indicate that rather than following standard gas-driven sublimation patterns, this exocomet is undergoing significant structural disintegration, shedding large-scale material and boulders as it recedes from perihelion.

# 6- SLC of Exocomet 3I/ATLAS

Using the methodology described in Section 2, we have created the phase plot of comet 3I/ATLAS, using data from the MPC, COBS and space data (Figure 7)

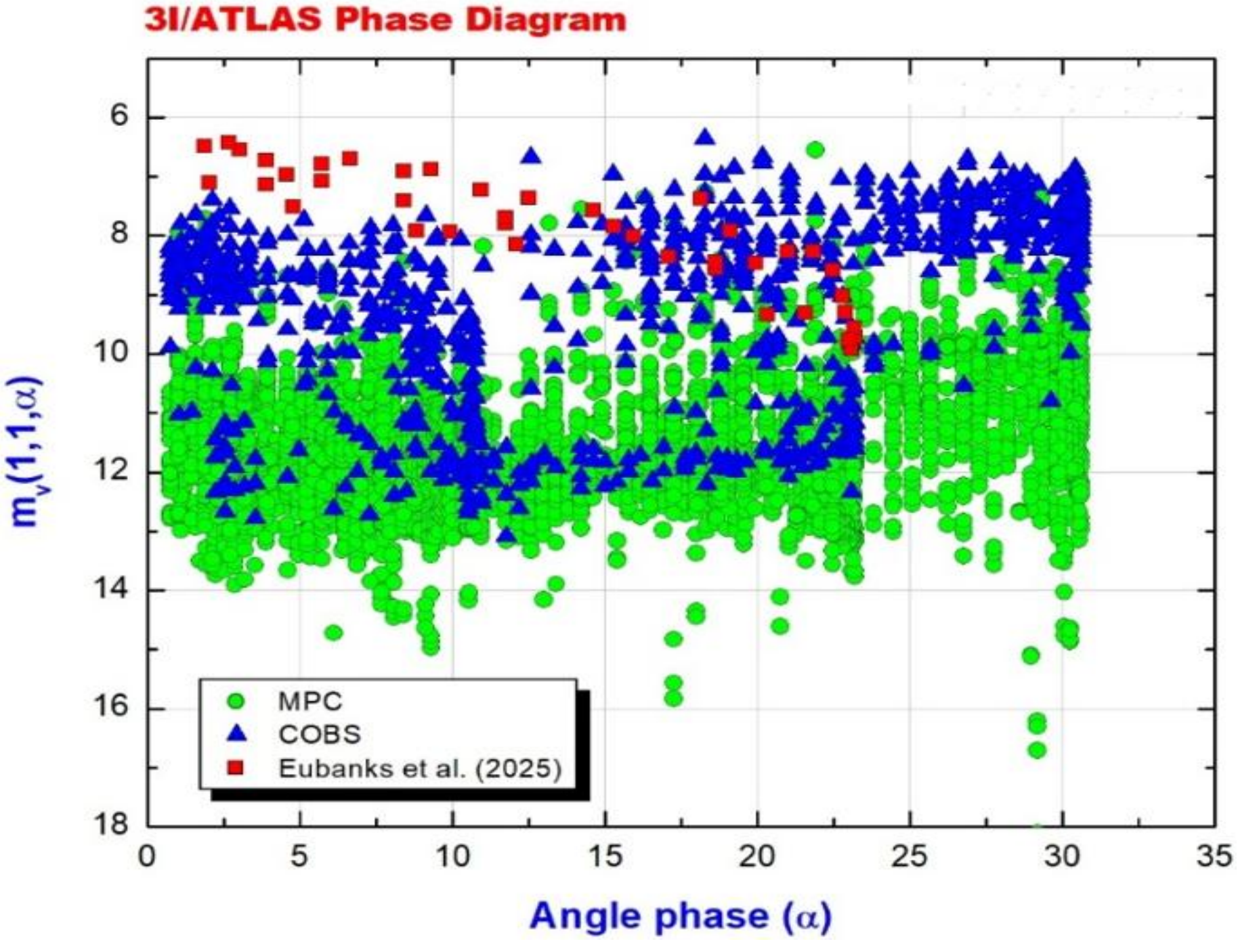

Figure 7. Phase Diagram of comet 3I/ATLAS. Each color represents a different data set. The diagram shows a chaotic distribution, indicating that there is no observed phase. That means that the surface cannot be seen, and there is an optically thick atmosphere surrounding the nucleus. If it had asteroidal behavior, there would be a single line with negative slope. This behavior implies that the object holds a coma during all its visible orbit.

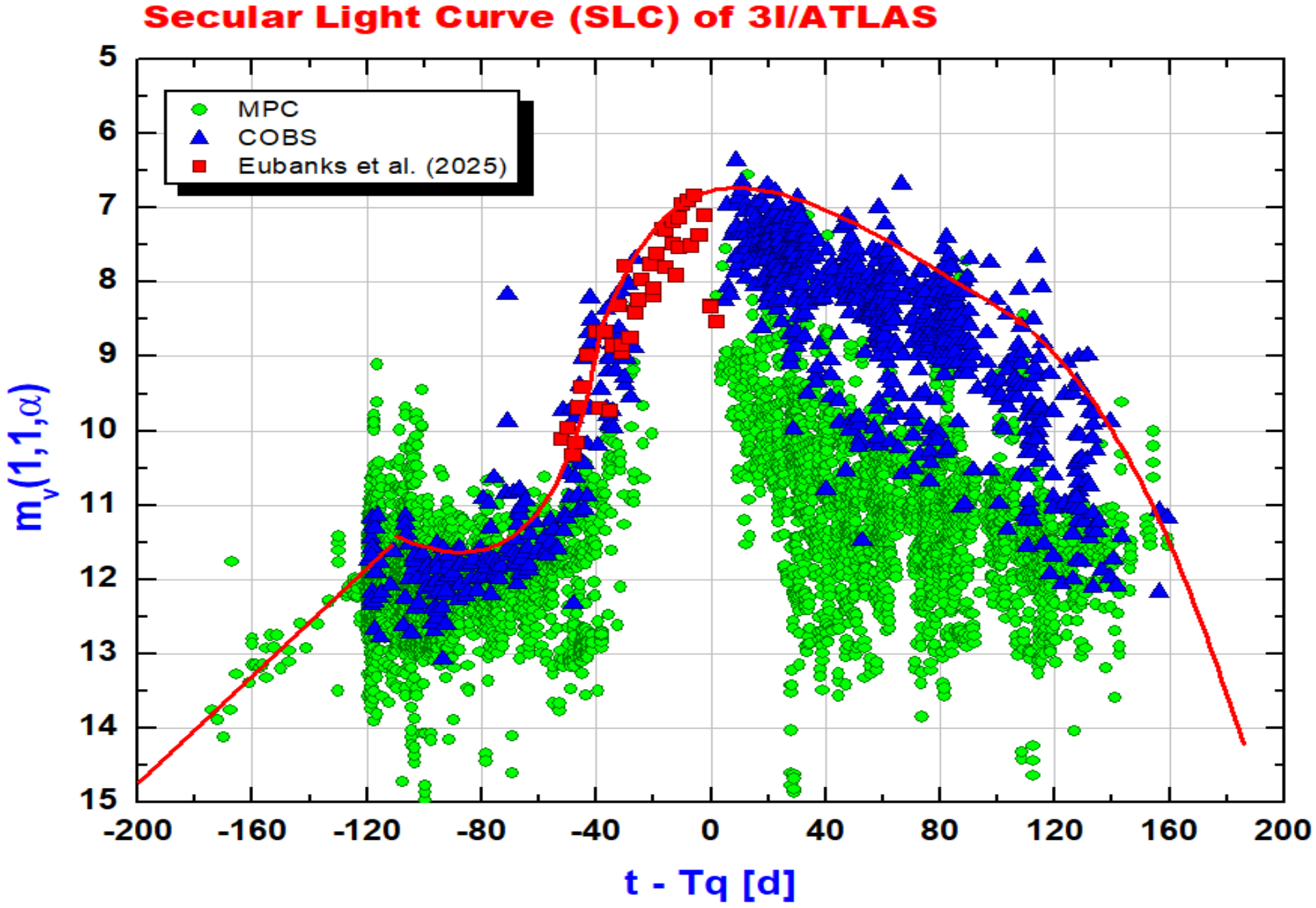

Figure 8. SLC time plot of comet 3I/ATLAS. The MPC observations show a much larger dispersion of the data than the COBS observations, and a systematic error downward. There is a significant increase in brightness starting at -45 days before perihelion that we interpreted as the onset of water sublimation. The maximum reduced magnitude reached by the comet was $m_V(max)=6.8\pm0.15$ mag. This plot and the next one may be considered as two of the highlights of the SLC Methodology. The other highlight is the Comet Evolutionary Diagram.

The advantage of using the time plot is that since time appears in the x-axis, the integral under the curve gives the budget of whatever is plotted. The following information can be extracted from Figure 8. First, there is a clear separation between the two datasets. In the right-hand side, the COBS observations are much brighter than the MPC observations. This is because COBS contains many visual observations, and the brain-eye combination can extract the whole flux from the coma. While the MPC show fainter magnitudes because the MPC dataset is an astrometric database, and astrometrist tend to use small apertures to determine the centroid. Second, there is a most estrange feature between -120 days and -45 days before perihelion that we call a photometric anomaly (Figure 9).

The maximum brightness did not take place exactly at perihelion but about 15 days later which implies the presence of a thermal lag of unknown origin. It could be due to a thick dust layer on the surface, or it could be due to a tilted rotational axis.

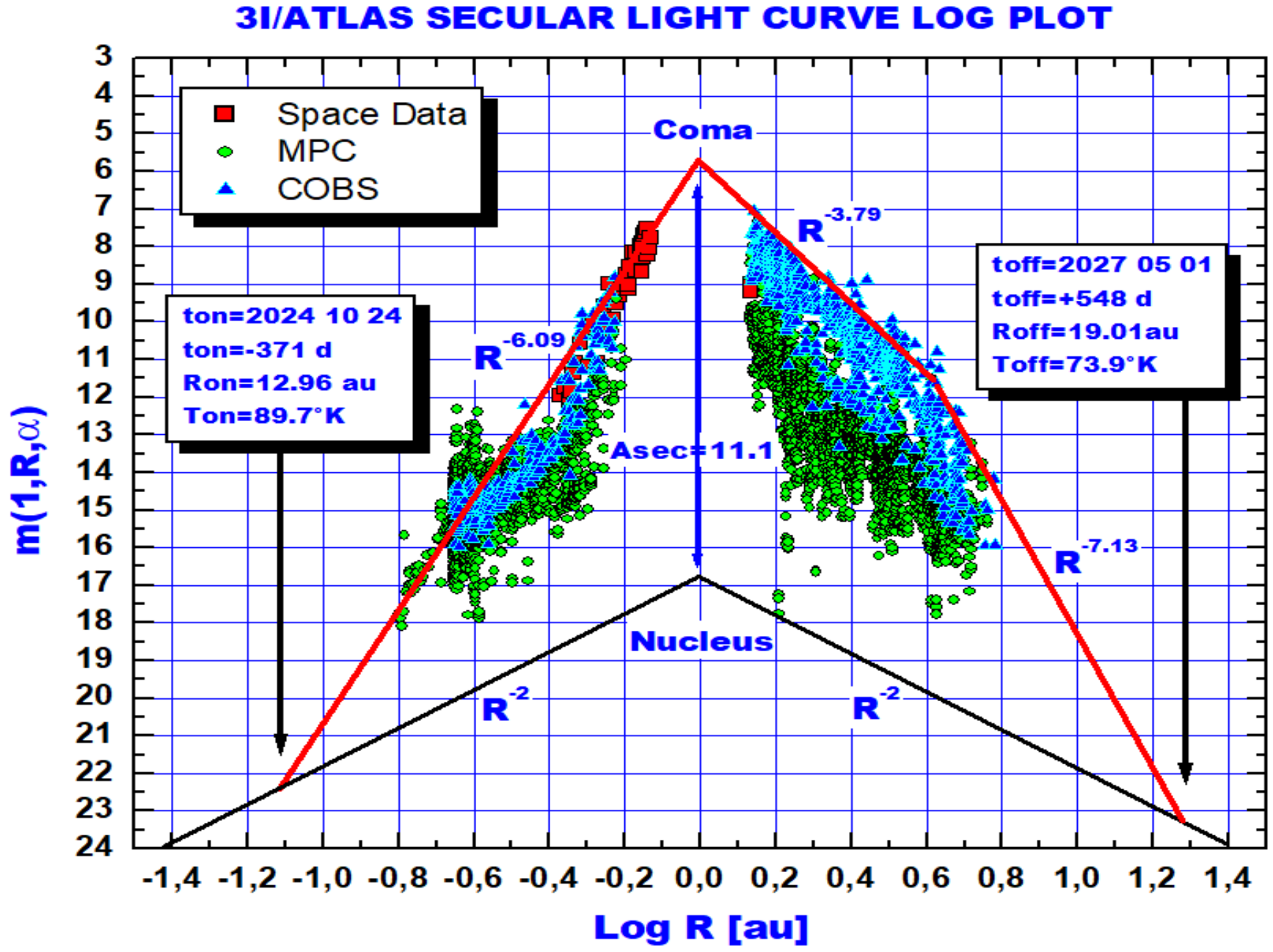

Figure 9. SLC Log Plot of 3I. This alternative way to analyze the data is part of the SLC Methodology. The horizontal axis is the Log R, the solar distance. The advantage of this plot is that powers of R plot as straight lines. So, it is easy to predict areas with no data. The data shows that the comet experienced a photometric anomaly between -120 days and -45 days, for a total of 75±5 days. This is not normal behavior for a comet. The absolute magnitude at the bottom in the form of a pyramid is the size derived by Man-To Hui et al. (2026) converted by us to absolute magnitude $m_V(1,1,\alpha)=16.9\pm0.3$ using an albedo pv=0.04. The turn on date was ~2024 10 24±10 d and the predicted turn off date ~2027 05 01±10 d.

Man Tui et al. (2026) calculated a nucleus radius of 1.4 km. We are interested in knowing to what absolute magnitude this value corresponds. To do that we will use the following equation given by Jewitt (1991) where pv is the geometric albedo, R the solar distance, Δ the Earth distance, $m_{sun}$ the solar magnitude and D the diameter of the nucleus in km:

$$p_V . r^2_{nuc} = 2.24x10^{22} R^2 \Delta^2 10^{[0.4(msun - mv(1,1,0))]} \quad (2)$$

which can be written in a more amicable form:

$$\text{Log} [ p_V D^2 / 4 ] = 5.654 - 0.4 m_v(1,1,0) \quad (3)$$

If the albedo pv=0.04 it can be simplified further:

$$\text{Log } D = 7.654 - 0.4 m_v(1,1,0) \quad (4)$$

This is the most amicable formula that you will find in the literature, to calculate the diameter from the absolute magnitude or the absolute magnitude from the diameter. There are more complicated ones.

We find an absolute magnitude of 16.9±0.3, and this value is plotted in Figure 9 in the lower part of the diagram as a pyramid. This procedure allows us to determine four parameters listed inside the plots: the turn on and turn off of the cometary activity ton= -371±10 d, while toff= +548±20 d, the dates of turn on and turn off, ton= 2024 10 14±10d and toff= 2027 05 01±20d. These correspond to solar distances Ron= -12.96±0.10 au and Roff=19.01±0.10 au, and temperatures Ton=89.7°±5°K and Toff=73.9°±5° K.

To calculate the temperature, we used the following calibration (Ferrín, 2013):

$$T[°K] = 323° \pm 5° / \mathrm{SQRT}(R)\ (au) \qquad (5)$$

The ratio Roff/Ron= 1.46 implies that sublimation takes place in depth, since this implies a lag time in activity.

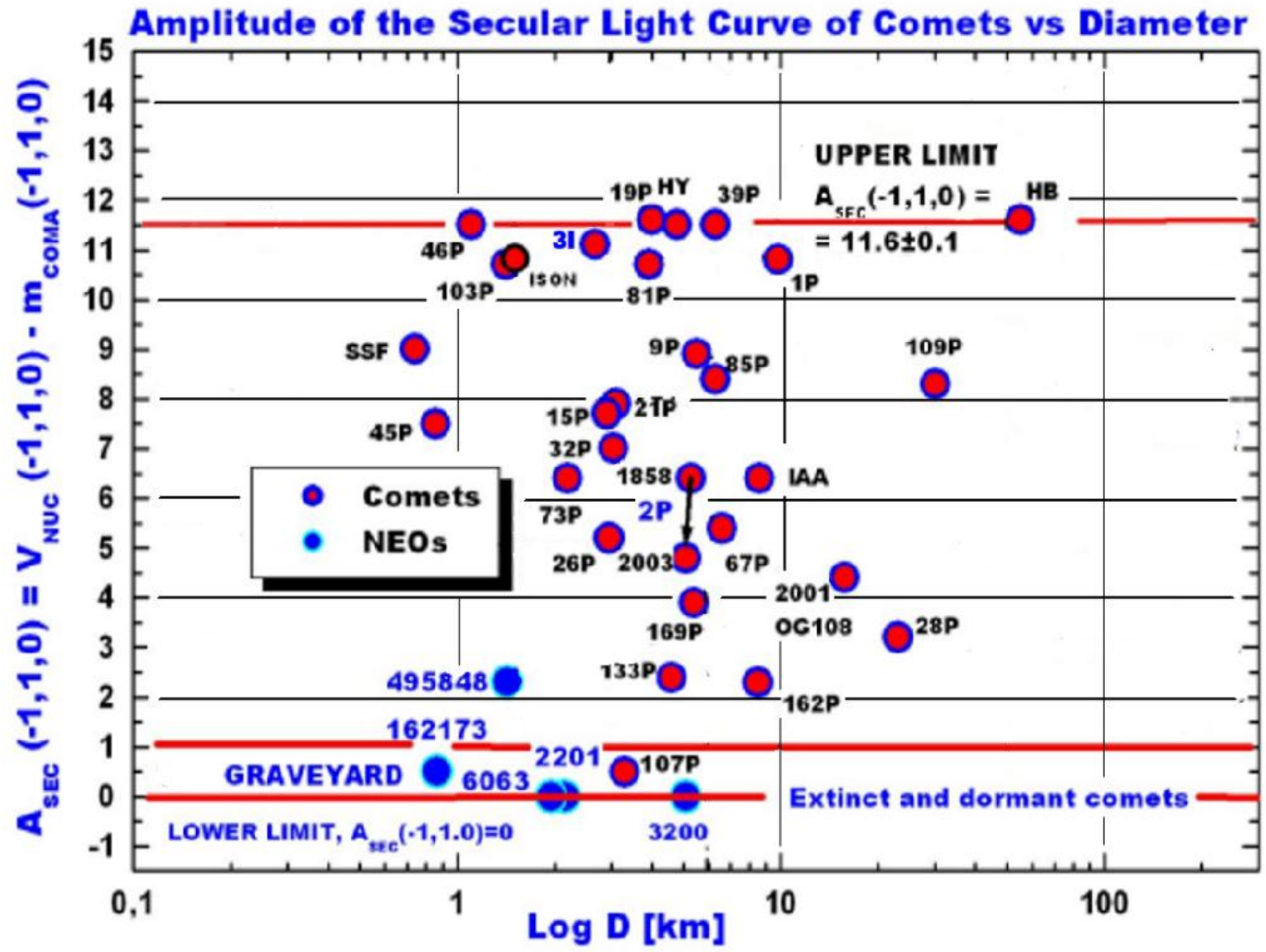

Figure 10. The amplitude of the Secular Light Curve at 1 au vs Log of the diameter. We learn that there is a maximum of activity Asec<11.6±0.1 mag for solar system comets. This is the maximum output that a comet can have, a physical limit. 3I/ATLAS is very near this limit with Asec=11.1±0.3 magnitudes.

Another important statistic that can be derived from Figure 9 is the amplitude of the activity at 1 au, Asec= 11.1±0.2 magnitudes. This value can be plotted in Figure 10 showing the Asec of several solar system comets, vs their diameter. We learn that solar system comets have a maximum activity of 11.6±0.1 magnitudes. Comets 46P, 19P, Hayakutake, 39P and Hale Bopp show maximum activity on the same line, implying the existence of an activity limit. Comet 3I/ATLAS has Asec=11.1±0.2 magnitudes close to this limit and implies that its activity is almost at its maximum possible.

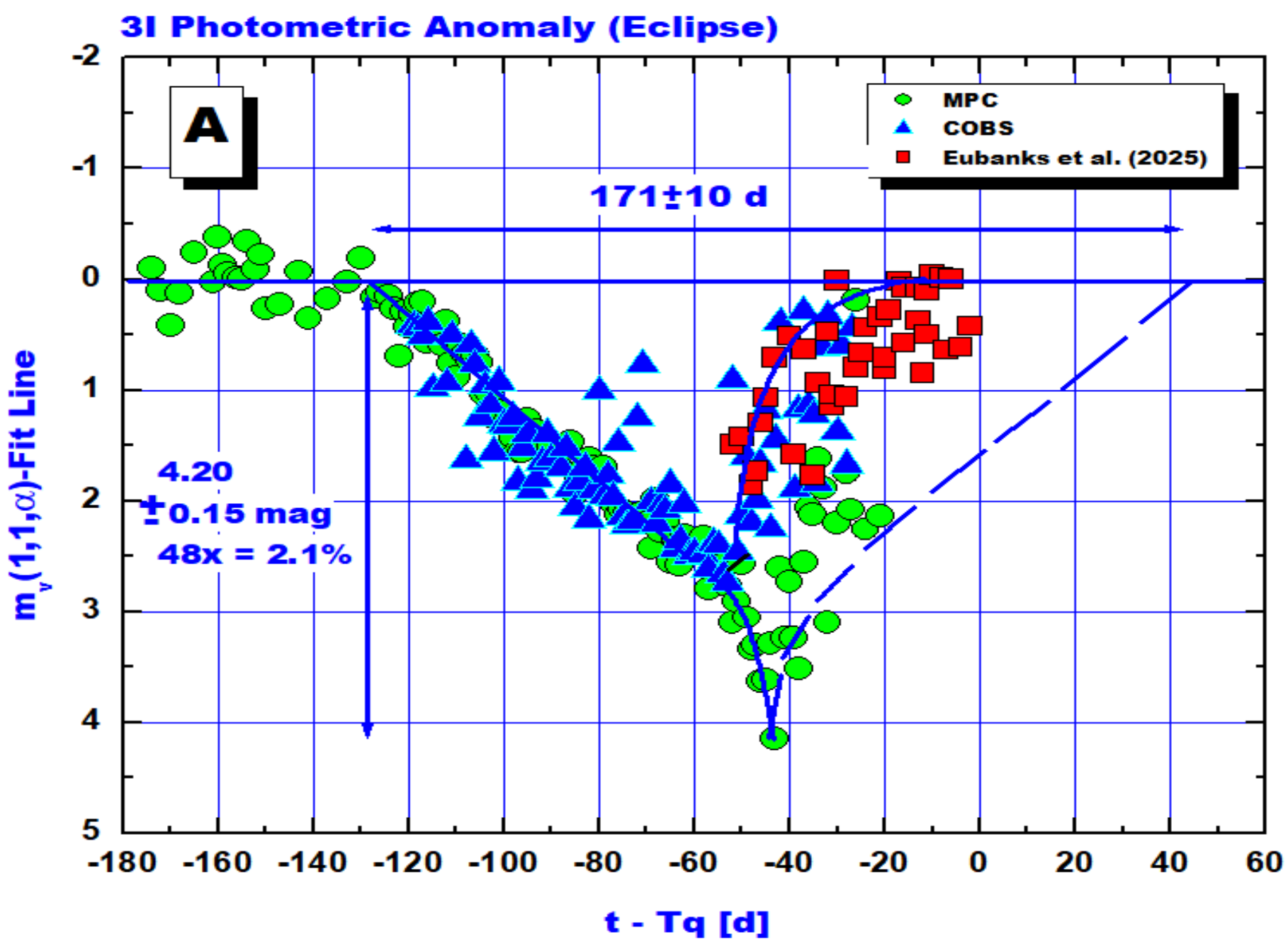

Figure 11. The envelope of the dataset has been subtracted from the observations, and now the curve is flat. Notice the agreement between the MPC and COBS. The photometric anomaly is clearly seen. The feature looks like an eclipse that has been interrupted by the onset of water sublimation. The conclusion is that there is evidence that the exocomet might be double. However, since we only have information on half of the eclipse the light curve is odd, the interpretation is complex and requires a detailed analysis beyond the scope of this paper.

## 8- Comet Aging Regimes: How comets die

Comets age, and there are three basic smooth models proposed to explain aging, excluding catastrophic disintegration (Figure 12).

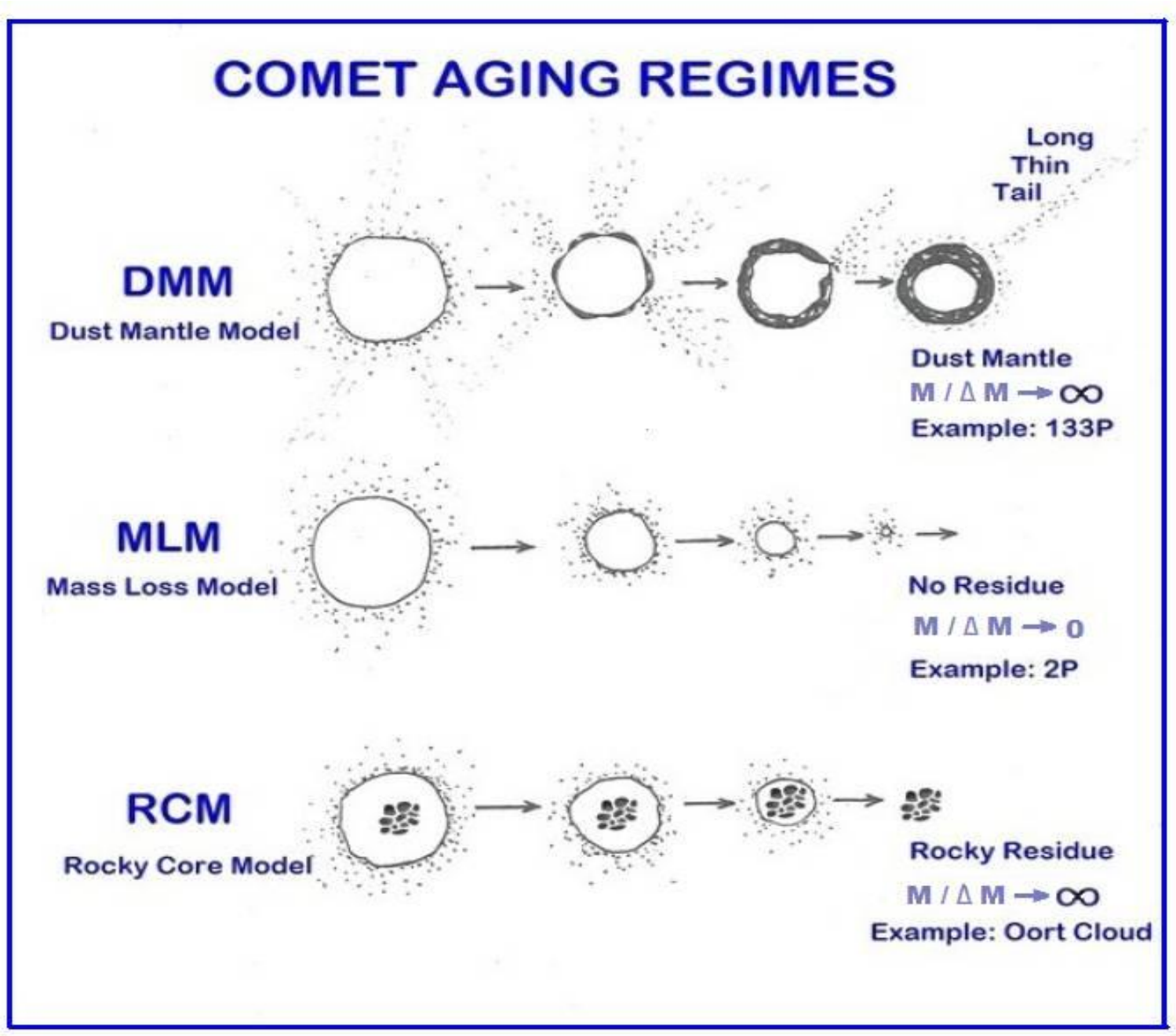

Figure 12. Comet Aging Regimes. We find that comets may lie in any of these three aging regimes, described in the text. Since M is the total mass of the object and $\Delta M$ is the mass lost in one apparition, the ratio $M/\Delta M$ measures the number of returns left, assuming a linear decay of $\Delta M$. Notice the evolution of the parameter $M/\Delta M$ vs time. In the first regime $M/\Delta M \rightarrow \infty$, in the second $M/\Delta M \rightarrow 0$ (the opposite direction), and the third to $\infty$ too. In this work the ratio $M/\Delta M$ is a very important parameter because it is the y-axis in the Comet Evolutionary Diagram (CED) studied later on. As a consequence, the Dust Mantle Model sends comets moving up, while the Mass Loss Model sends comets moving down. Thus, the two groups can be separated neatly.

The Dust Mantle Model (DMM) represents a comet composed primarily of ice and dust, with no rocky nucleus. Disintegration occurs slowly as the particles that make up the outer dust mantle escape from the surface. However, maybe because the nucleus is small or the dust is large, the dust falls back on the nucleus creating a deep dust mantle that quenches the sublimation, until the layer is so deep that no sublimation takes place. The nucleus becomes a completely masked comet and hides as an asteroid. Then, $M/\Delta M \rightarrow \infty$.

The second model, called the Mass Loss Model, applies to a comet that sublimates the ice and no dust is retained on the surface. The comet diminishes in size systematically until it loses all its ice and sublimates away leaving no trace. Then $M/\Delta M \rightarrow 0$.

Finally, the Rocky Core Model (RCM) corresponds to a comet consisting of a core composed of a mass of rocks of certain density, embedded in a matrix of glue-ices and surrounded by a layer of ice and dust. Then $M/\Delta M \rightarrow \infty$. These tendencies will be important when we reach the Comet Evolutionary Diagram, CED, because the first group will move up in the diagram, while the second group will move down. That precisely is one of the nice properties of the CED: it separates the two groups neatly.

## 9- Comet Evolutionary Diagram (CED): The family tree

In 2005 the first author raised the question of why comets didn´t have an evolutionary diagram as stars do. It took him 5 years to find the two parameters needed in the horizontal and vertical axis, since luminosity and temperature were not valid for comets. Their brightness and temperatures are the result of solar insolation as they do not emit light. It was found that the axis labels were ML-AGE = Mass Loss Age in the x-axis, and the RR=Remaining Returns (=$r/\Delta r$, were r=radius of the object) in the vertical axis. The Mass Loss Age is the inverse of the total mass loss as a proxy for age, since comets diminish in size as they age. The first paper on this issue was published (Ferrín 2012), and subsequently 3 additional papers were published (Ferrin et al. 2013, 2014, 2019). So, this is the fifth paper on the CED.

It is the objective of this research line, to plot the three exocomets on the CED, to get insights into their physical nature and to compare them with solar system comets. 1I/Oumauma and 2I/Borisov will be considered in the next paper of this series. Here we will be concerned only with 3I/ATLAS.

## 9.1 Object size

Before we can plot the object in the CED, it is a necessary condition to have the size of the object, the SLC, and the total production rates of dust, H2O, CO2 and CO over all the orbit (Lisse et al, 2025a, 2025b, 2026a, 2026b). The comet is well past perihelion, so it is possible to make a reasonable assessment of these quantities.

Several authors have estimated the size, but the best determination is due to Man-To Hui et al. (2026) who obtained a radius $r = 1.3\pm0.2$ km using nucleus extraction and $r = 1.5\pm0.1$ km using nongravitational effects. Thus, we will adopt a mean radius of $1.4\pm0.15$ km, or an equivalent diameter of $D = 2.8\pm0.3$ km.

## 9.2 Calculation of the Dust Mass Loss

To calculate the Dust Mass Loss ($Af\rho$), we will use the equation:

$$Q(\text{Dust}) = \sum_{\text{Ton}}^{\text{Toff}} Af\rho(t)\, dt \qquad (6)$$

To determine the amount of dust, we will use the Afρ-parameter defined by A'Hearn et al. (1995) that comes expressed in cm. On March 20th, 2009, we wrote to Mike A´Hearn and asked him what the conversion factor to kg would be and this was his answer (quote): *"If I want an interpretation I use another empirical correlation that was worked out by Claude Arpigny and which is discussed in our 1995 paper on the photometry database* (A'Hearn et al., 1995). *That relation as stated there is that an Afρ of 1000 cm is equivalent to a mass loss rate of one metric ton per second. Equivalently this means that an Afρ of 1 cm is equivalent to a mass loss rate of one kg per second. This relationship, of course, ignores differences in particle size distribution, differences in outflow velocity, and lots of other things, but it is useful in getting an understanding of patterns as long as it is not overinterpreted"* (private communication).

Our first dataset is available at the Cometas_Obs (Astrosurf) (2026) site, which covers 240 days of observations, and it is homogeneous. Scarmato (2025), Jewitt and Luu (2025) and Fraser et al. (2026) also made measurements of Afrho, and their datasets are plotted in Figure 13. The dust behavior can be described by two linear production rates. We believe that the most probable reason for the scatter is the measuring aperture. It is a common recurrent error in the literature, to try to extract the whole flux of the coma with small apertures. This produces downward errors. That is why we adopt the largest measurements because they most probably capture a total coma contribution. The datasets of Fraser et al. (2026) and Trigo et al. (2025) lie above the other datasets, so we take them as reference. We had the same problem when we considered the visual and CCD observations. We found a Total Mass Loss (Dust) = $8.9\times10^{10}$ kg.

Fraser et al. (2026) mention in their work, that the calculation of the mass-loss rate of dust can be done under simplified assumptions, like grain radius, ejection velocity and grain density, but that this calculation is an order of magnitude estimate, since it is inherently speculative and highly sensitive to the adopted grain properties, with the result varying by at least an order of magnitude (Gillan et al., 2025). Thus, they suggest that the resulting values should be regarded as approximate. Their arguments may explain why in the subsequent plots, different datasets may differ by large amounts and reconciling them is not that simple. In conclusion, their argument is not only valid for the dust, but extends to the water, $CO_2$ and CO production rates determinations too.

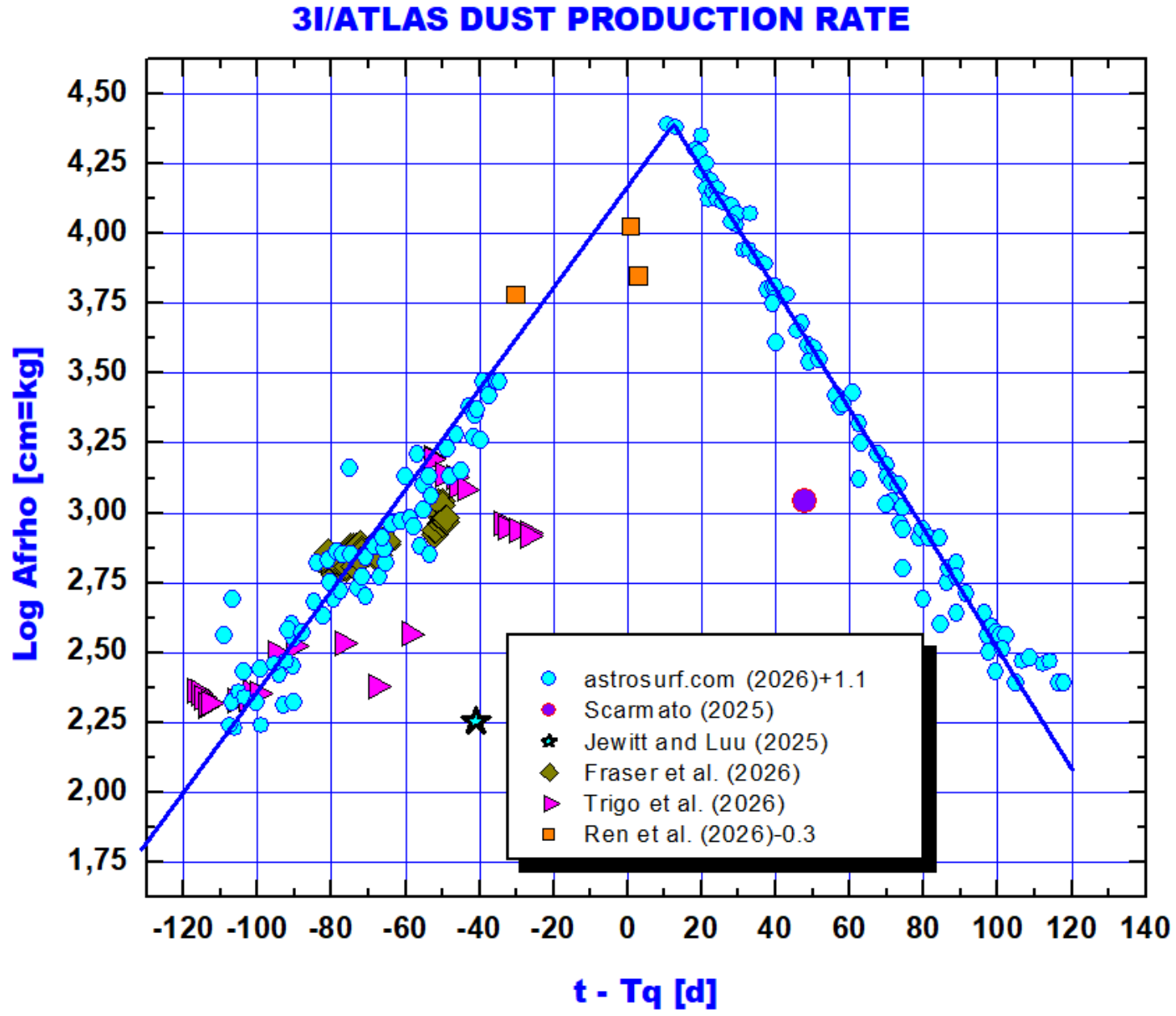

Figure 13. Dust production rate vs time with respect to perihelion. The integral below the two lines gives the Total Mass Loss (Dust) = $8.90 \times 10^{10}$ kg. This plot exhibits the difficulty of measuring the production rate of this species. The discrepancy may reach an order of magnitude. We believe that part of this discrepancy lies in the size of the measuring aperture that is too small to extract the whole flux from the coma, a recurrent error in the literature.

## 9.3 Water Production Rate

At about -45 d before perihelion, the SLC shows a steep increase in magnitude. We believe this uptrend is due to the onset of water ice sublimation. To verify this hypothesis, we extracted the water turn on of several comets from the Atlas of Secular Light Curves (Ferrín, 2010). The histogram of turn on points is shown in Figure 14. We see that the turn on of 3I at -2.5 AU from perihelion, is completely consistent with other turns on distances of other solar system comets. Thus, the turn on point of 3I is consistent with water ice sublimation.

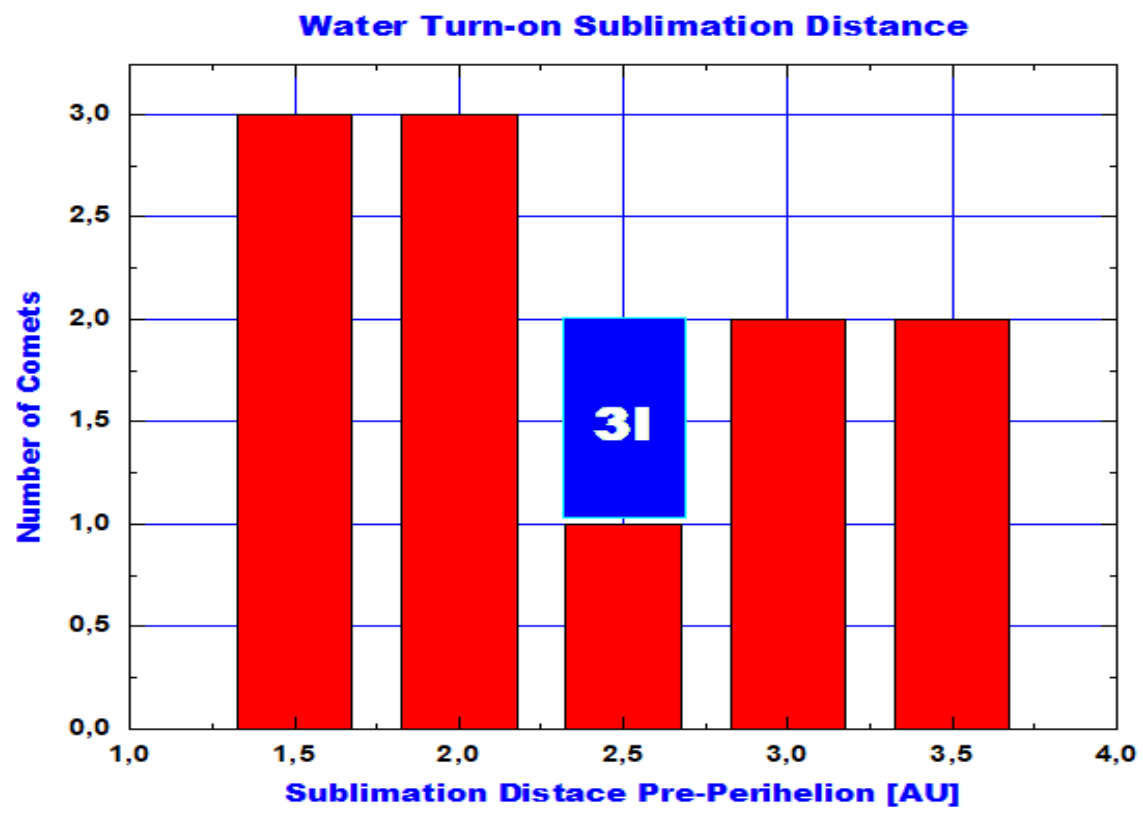

Figure 14. A histogram of turn on distances of water (H2O), extracted from the Atlas (Ferrín, 2010).

In Figure 15 we show the water production rate in comet C/2009 P1 Garradd, near 2 au pre-perihelion, as a function of aperture size (Combi et al. 2013). We see that only the largest apertures are capable of extracting the flux from the whole coma. The flux increases asymptotically as the aperture increases, allowing the definition of infinite aperture water production rates. This is one reason why it is advisable to adopt the largest values of the water production rate measurements as the correct interpretation of the data. This reasoning applies to any production rate, and to the visual and CCD magnitudes.

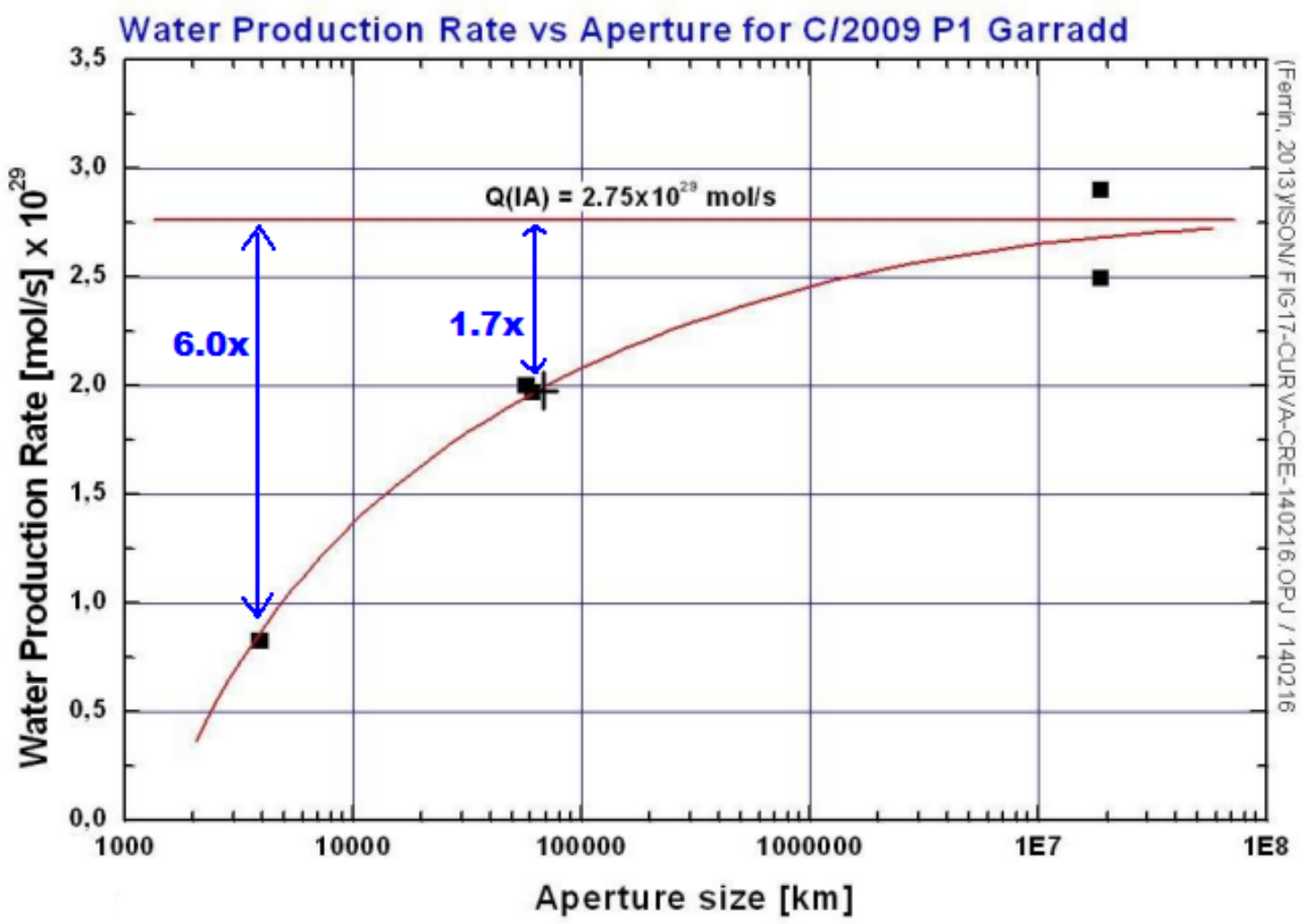

Figure 15. The data of Combi et al. (2013) shows that smaller apertures capture smaller amounts of gas. The best value is the infinite aperture production rate. The error is downward. This same problem is exhibited by all the other species and even the observed crude magnitudes.

To calculate the total mass lost by the comet we need the mass loss by water, one of the most abundant components. To do so, we compiled all the datasets available in the literature, Combi et al. (2026), Hanjie

Tan et al (2026), Li et al. (2025), Hutsemekers et al (2025), Xing et al. (2025), Biver et al. (2025). The Combi et al. (2026) dataset lies a factor of ≈5x above the other datasets, and the dataset of Hutsemekers et al (2025) lies about ≈12x below the other datasets. Both authors mention difficulties in the water calibration. So, we had to apply corrections to make sense of the water production rate of this exocomet. We applied a 3rd degree polynomial to this dataset to be able to calculate the production rate at any date. The result of this calculation appears in Figure 16.

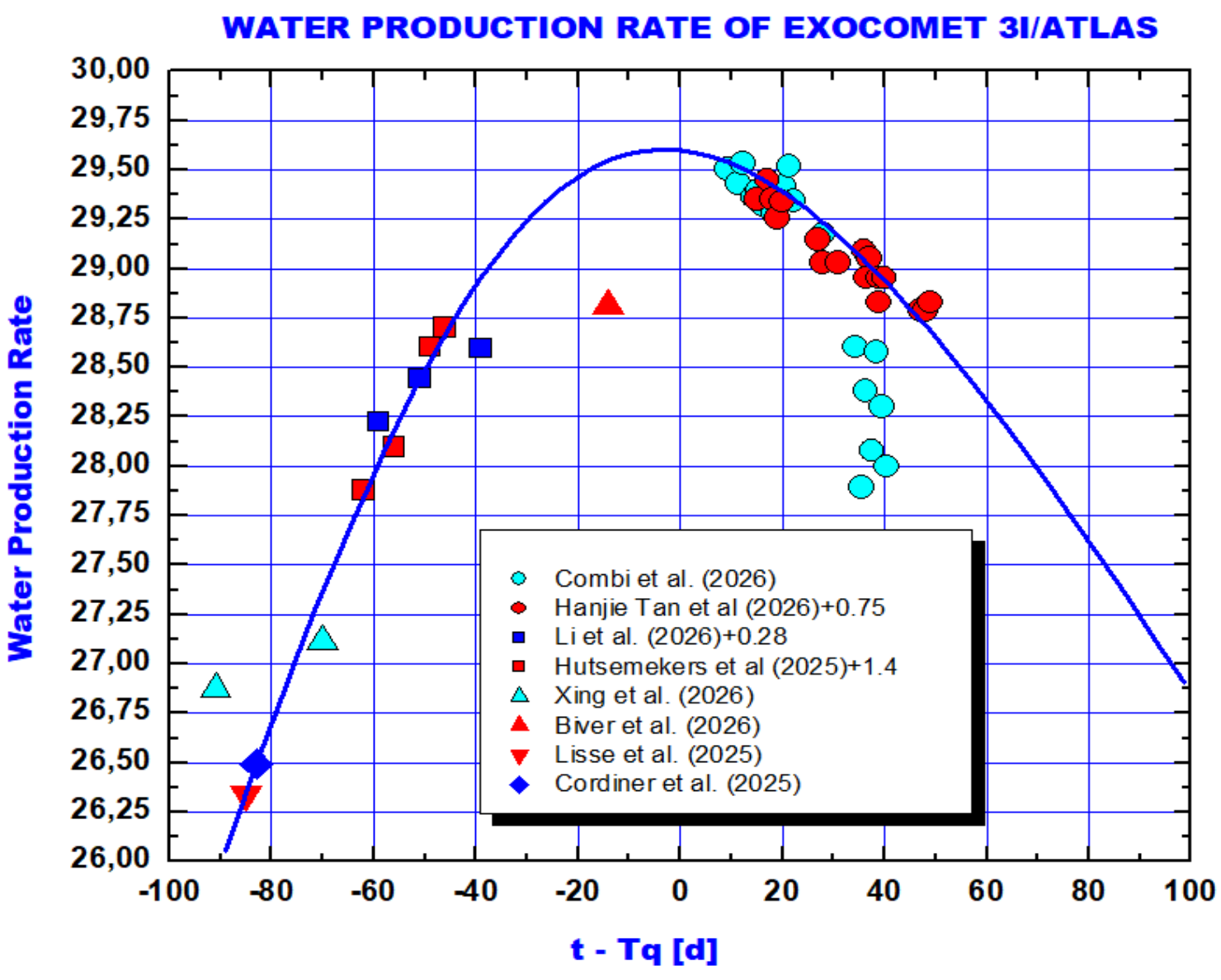

Figure 16. The water production rate of 3I/ATLAS as a function of time. The integral below the fit is the total water mass loss and equals Qtotal(H2O) = $6.05 \times 10^{10}$ kg.

## 9.4 CO2 Production Rate

CO2 is another important component of comets. We are interested in determining the total production of CO2 in this exocomet. The result of the literature search can be seen in Figure 17. Since only 5 data points were available, we fitted the same production rate vs time for water, and the fit looks quite good, so we proceeded to integrate it below the curve.

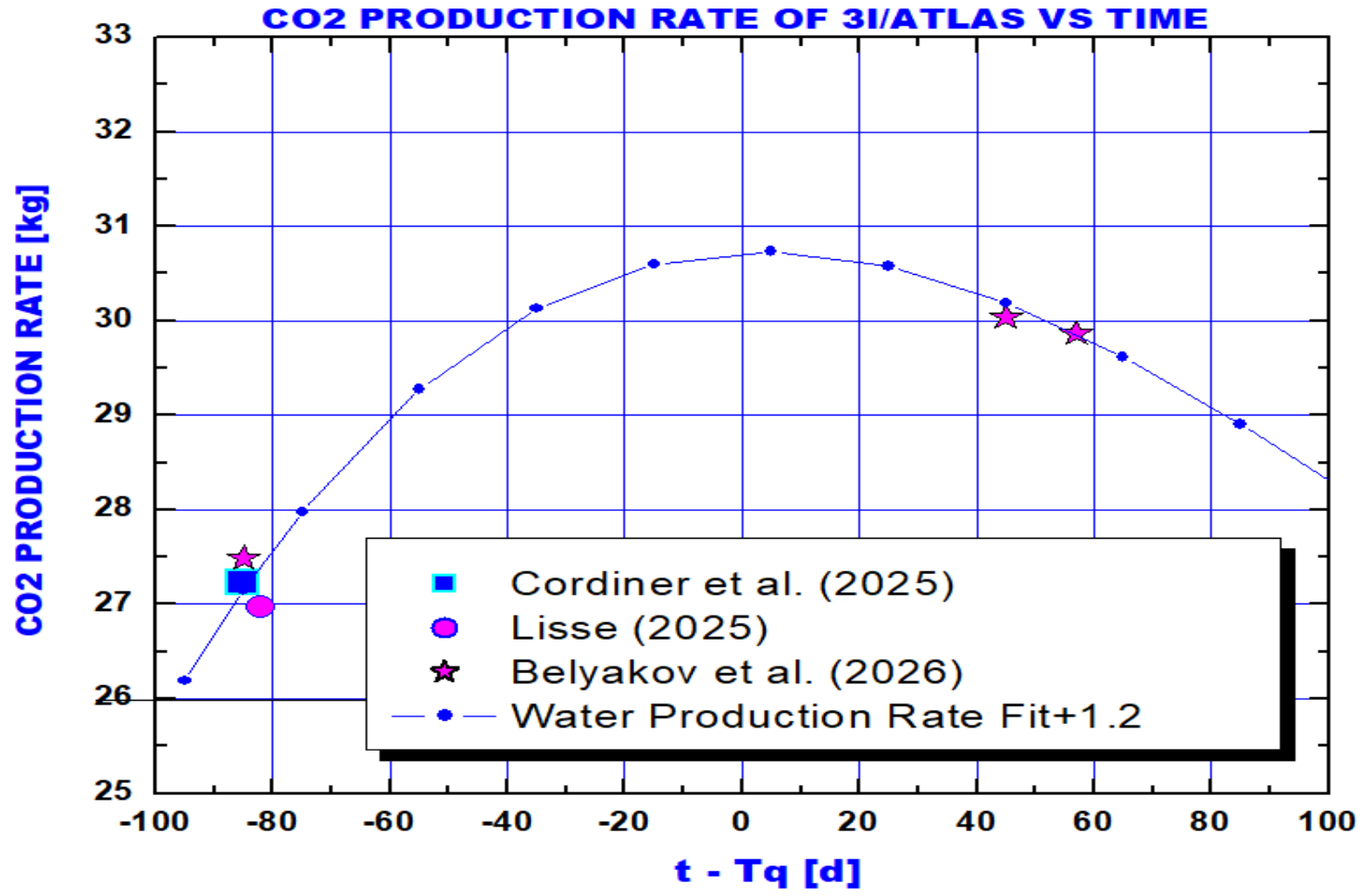

Figure 17. The CO2 production rate of 3I/ATLAS as a function of time. The integral below the fit is the total CO2 mass loss, and equals Qtotal(CO2) = $1.06X10^{12}$ kg.

## 9.5 CO Production Rate

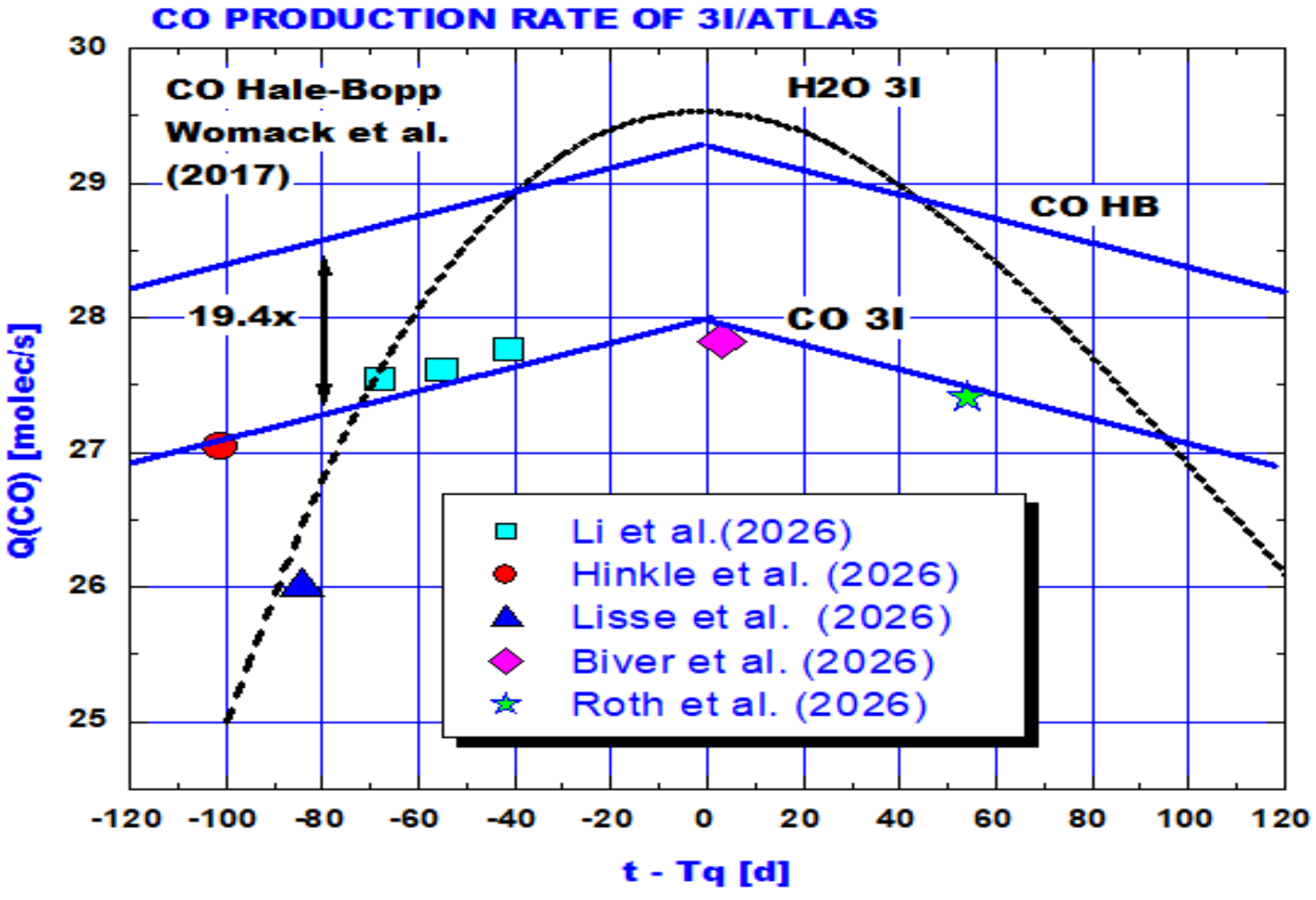

Figure 18. The CO production rate of comet Hale-Bopp, and 3I measured by several authors. The dashed line is the H2O production rate of 3I for comparison. The Li et al. (2006) values lies a factor of 19.4 below the HB production rate. HB follows a law Q(CO) = $3.5 \times 10^{29} / R^{-2}$ while 3I follows law $2.5 \times 10^{28} / R^{-2}$ . Total Q(CO) = $4.6 \times 10^{9}$ kg. Notice that this dataset did not require any correction.

Zhao, R. et al (2026) conclude that Q(CO)/Q(H2O) $\geq 7.5$ at 1.9 au. At $3.84 < R < 3.94$, Hinkle et al. (2026) find an upper limit $<1.1x10^{27}$ mol/s. While Lisse et al. (2026a) derives a CO production rate of $1.0x10^{26}$ mol/s at 3.3 au or -84 d pre-perihelion. This shows the difficulty of measuring these quantities by authors that use different methods and different apertures, and issue that was considered above.

Womack et al. (2017) and Gunnarsson et al. (2003) found that the CO production rate of comet Hale-Bopp depended on distance to the sun as $R^{-2}$. Womack et al. (2017) even gives an equation for this dependence: Q(CO) = $3.5 \times 10^{29} / R^{-2}$. We find that the equation for 3I is Q/CO) =$2.5 \times 10^{28}/R^{-2}$ . The ratio is Q(CO)(HB) / Q(CO)(3I) = 19.4.

## 9.6 Nucleus Mass

The calculation of the object mass is a requisite to create the CED. We will adopt from Man-To Hui et al. (2026) a mean radius of 1.4±0.15 km. Because this exocomet is made of 93% $CO_2$ we will adopt the density of "dry ice" $\rho = 1560$ kg/ $m^3$.

$$M(nuc) = (4/3)\, \pi\, \rho\, r^3 \qquad (7)$$

For exocomet 3I we find M(nuc) = $1.78\times10^{13}$ kg

## 9.7 Total Mass Loss

The total mass loss can be calculated using the total mass loss per component:

$$\Delta M = \Delta M(Dust) + \Delta M(H_2O) + \Delta M(CO_2) + \Delta M(CO) \qquad (8)$$

$$= 8.90\times10^{10} + 6.05\times10^{10} + 2.12\times10^{12} + 4.6\times10^{9}$$

$$\Delta M = 2.274\times10^{12} \text{ kg}$$

Table 2. Percentage composition of substances in the nucleus.

| 3I (This Work) | %Dust= 3.9% | %H2O=2.66% | %CO2=93.23% | %CO=0.20% |
|---|---|---|---|---|

Since these calculations have been made from actual observations reported in the literature, our results are very robust. The conclusion is obvious, this exocomet is by far made of $CO_2$. This raises two questions: what is the density of $CO_2$ ice, and what is its albedo. For the density we will adopt the density of dry ice, 1560 kg/$m^3$. Since 93.23% is $CO_2$ and there is so little dust, the albedo might be much higher and it is unknown, so we will continue with the standard value 0.04 due to our ignorance.

## 9.8 Calculation of the shell depth Δr

At every return a comet loses a layer of mass of depth Δr:

$$\Delta r = \Delta M / 4 \pi \rho r^2 \quad (9)$$

For exocomet 3I we find Δr = 59 m, a large number.

## 9.9 Calculation of RR=r/Δr

The calculation of the remaining returns RR, can be performed using r/Δr or using 3 M/ΔM. They are the same.

$$RR = 3 M/\Delta M = 3 (4/3) \pi r^3 \rho / \Delta r 4 \pi r^2 \rho = r / \Delta r \quad (10)$$

For exocomet 3I, RR = r/Δr = 1400 m/59 m = 24 returns.

## 9.10 Mass Loss Age Calculation

The mass loss age has been defined as the inverse of the total mass loss as a proxy for age (Ferrín, 2010). The constant is the total mass loss experienced by comet 28P/Neujmin, to which a comet age of 100 years has been assigned. Comet ages have been calibrated to human ages.

$$\text{ML-Age} = 3.58 \times 10^{11} / \Delta M \quad (11)$$

For comet 3I, ML-Age(3I) = 0.16 cy, a baby comet.

## 9.11 Ratios

Many ratios have been calculated in the literature but all of them refer to specific time-intervals and show departures due to the aperture error. Our ratios have been calculated using the total mass loss and thus should be much more robust than individual determinations.

Table 3. Ratios measured from total mass loss.

| Ratios | From total mass loss |
|---|---|
| ΔM(Dust)/ΔM(H2O) | 1.47 |
| ΔM(CO2)/ΔM(H2O) | 35 |
| ΔM(CO)/ΔM(H2O) | 0.076 |
| ΔM(CO)/ΔM(CO2) | 0.0022 |
| ΔM(CO2)/ΔM(CO) | 461 |

In solar system objects, the mean dust/water ratio is about ~1. In poor dust comets the ratio may be ~0.1. Rotundi et al. (2025) found dust-to-gas ratios of ~4±2 for 67P. Fulle et al. (2017) report ratios varying from ~5-10. A'Hearn et al. (1995) show large diversity but supports order-unity dust to gas ratios. Mumma and Charnley (2011) conclude that dust-to-gas ≈1 is a representative average across comets. Thus, a value of 1.4 for 3I is reasonable.

These results imply that 3I/ATLAS is depleted in CO and enriched in CO2. 3I/ATLAS has a CO2 dominated coma, which may be explained partly by the preferential loss of CO over millennia because of its extreme volatility.

## 9.12 Ratios Displayed

Harrington Pinto et al. (2022) published a very nice paper, studying the production rate of H2O, CO2 and CO of 25 comets. Their plots are very interesting because they give a birds' eye view of the behavior of these species in the solar system. We will plot our values in an adapted copy of their plots (Figures 19, 20, 21).

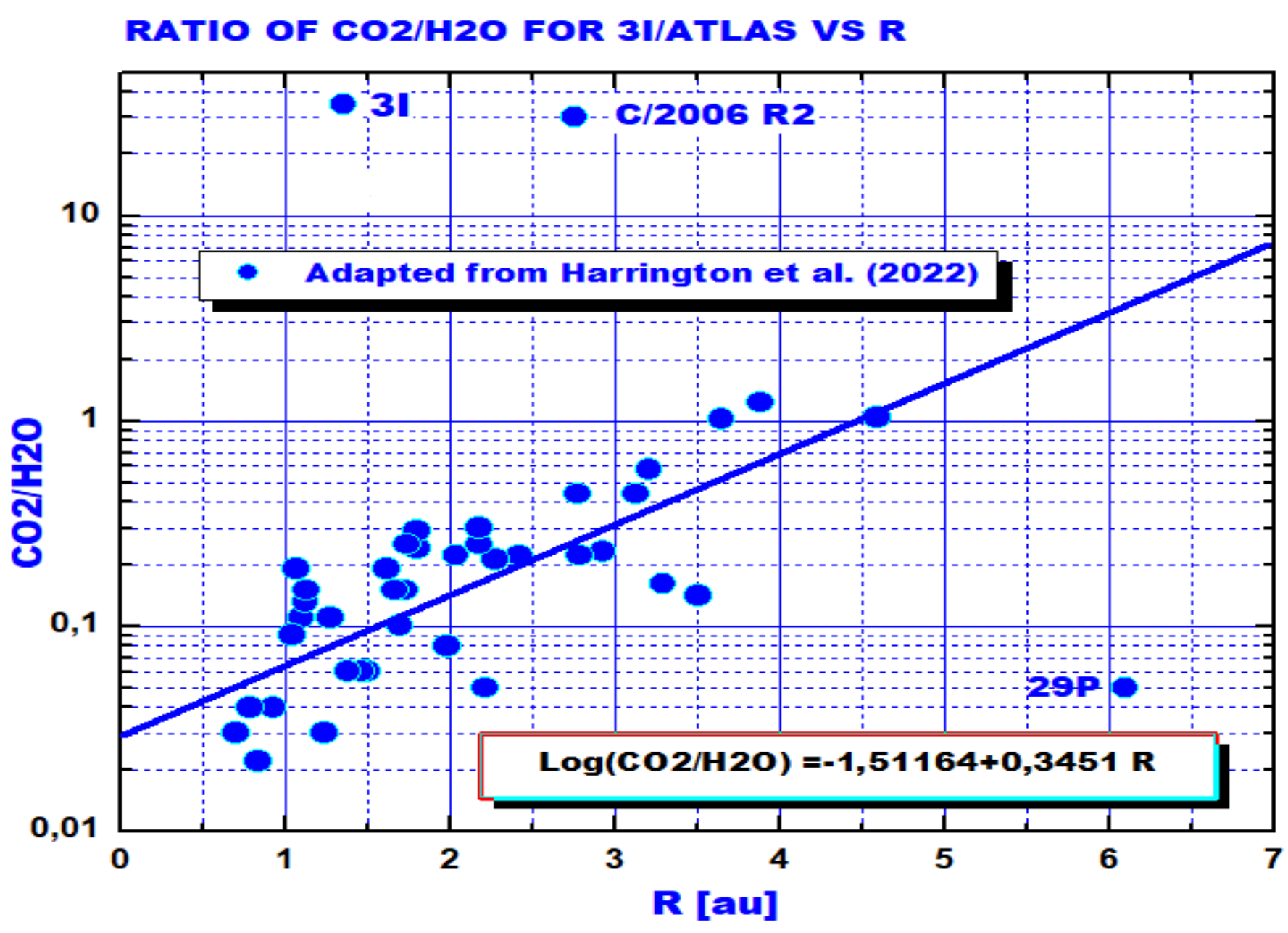

Figure 19. In this plot adapted from Harrinton Pinto et al. (2022) we see that 3I departs completely from the 40 data points of solar system comets. However, it has company from C/2006 R2, suggesting that it might be an Oort Cloud-type comet but from another stellar system.

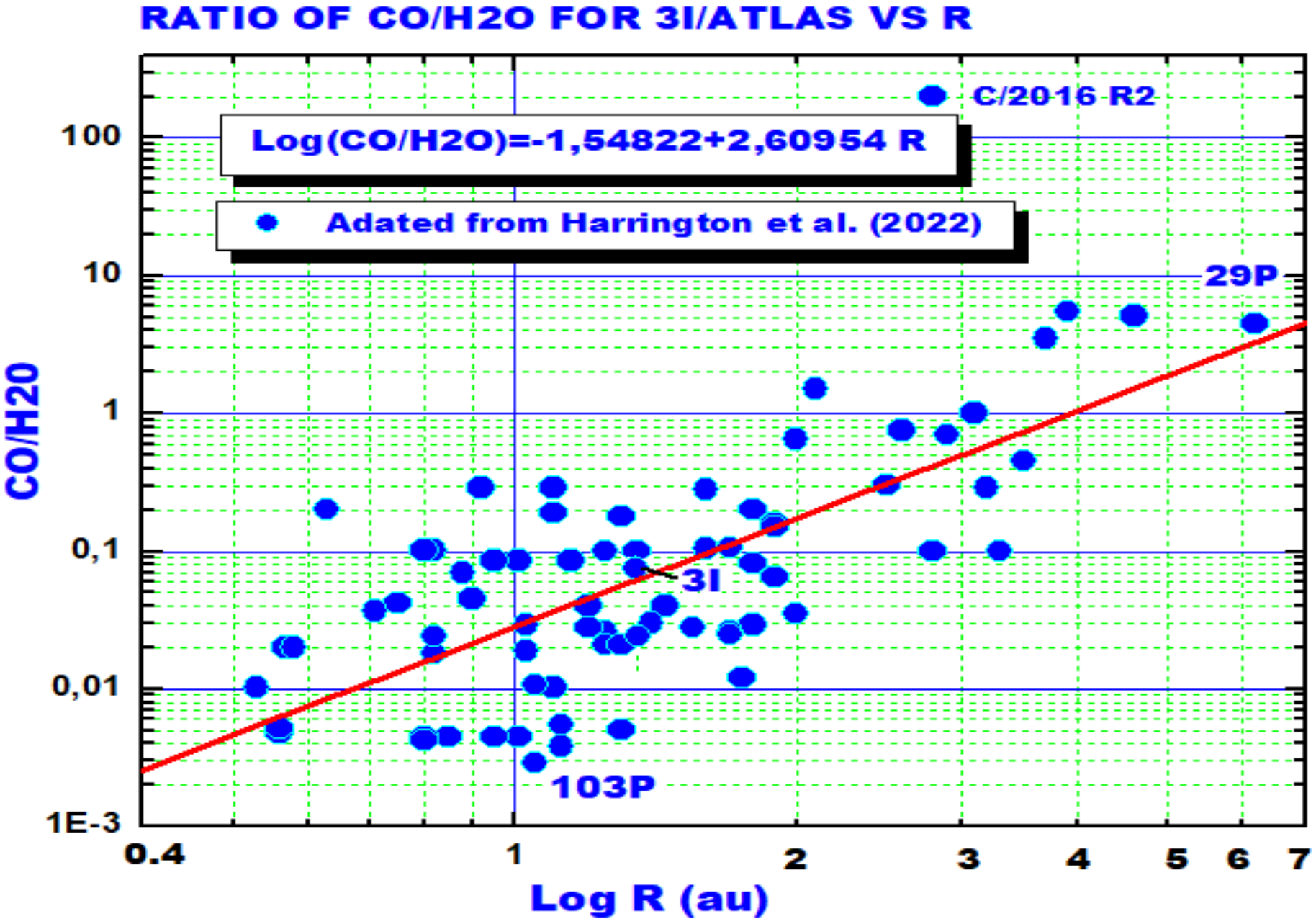

Figure 20. In this plot adapted from Harrinton et al. (2022) we see that exocomet 3I agrees with solar system comets very well.

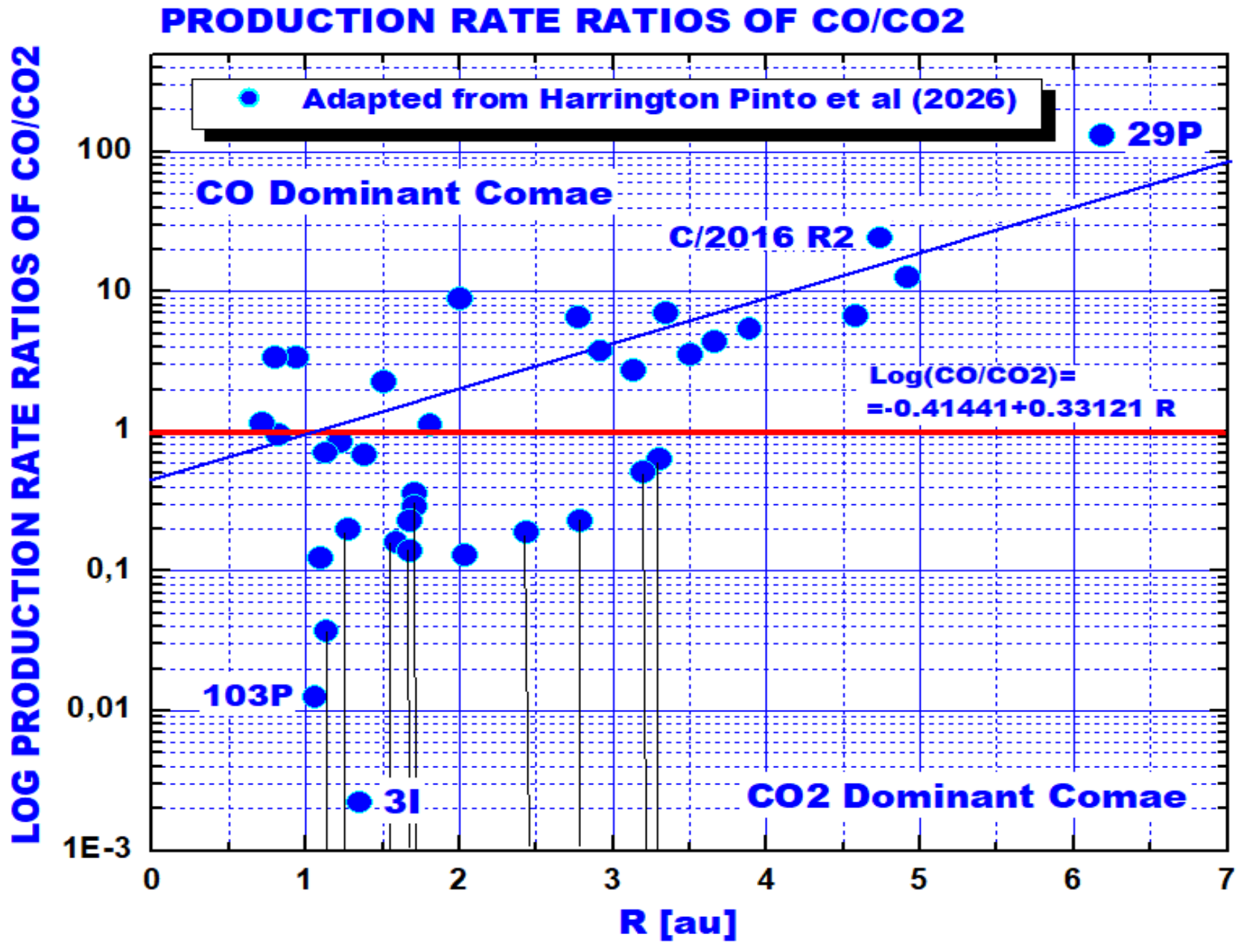

Figure 21. Harrigton Pinto et al. (2022) used 36 data points, to calculate the distribution of solar system comets in these parameters space. The diagram divides the plot into two regions. Above CO comets dominate, and extreme members are 29P and C/2016 R2, implying that CO dominated comets exist in our solar system and in the Oort Cloud. However, 3I and 103P are dominated by $CO_2$, and it is clear that $CO_2$ dominated comets exist in our solar system and in the stellar system of 3I. Then, 29P, C/2016 R2, 103P and 3I are extreme comets compositionally. (Adapted from Harrington Pinto et al. (2026).

## 9.13 Understanding the Evolutionary Diagram

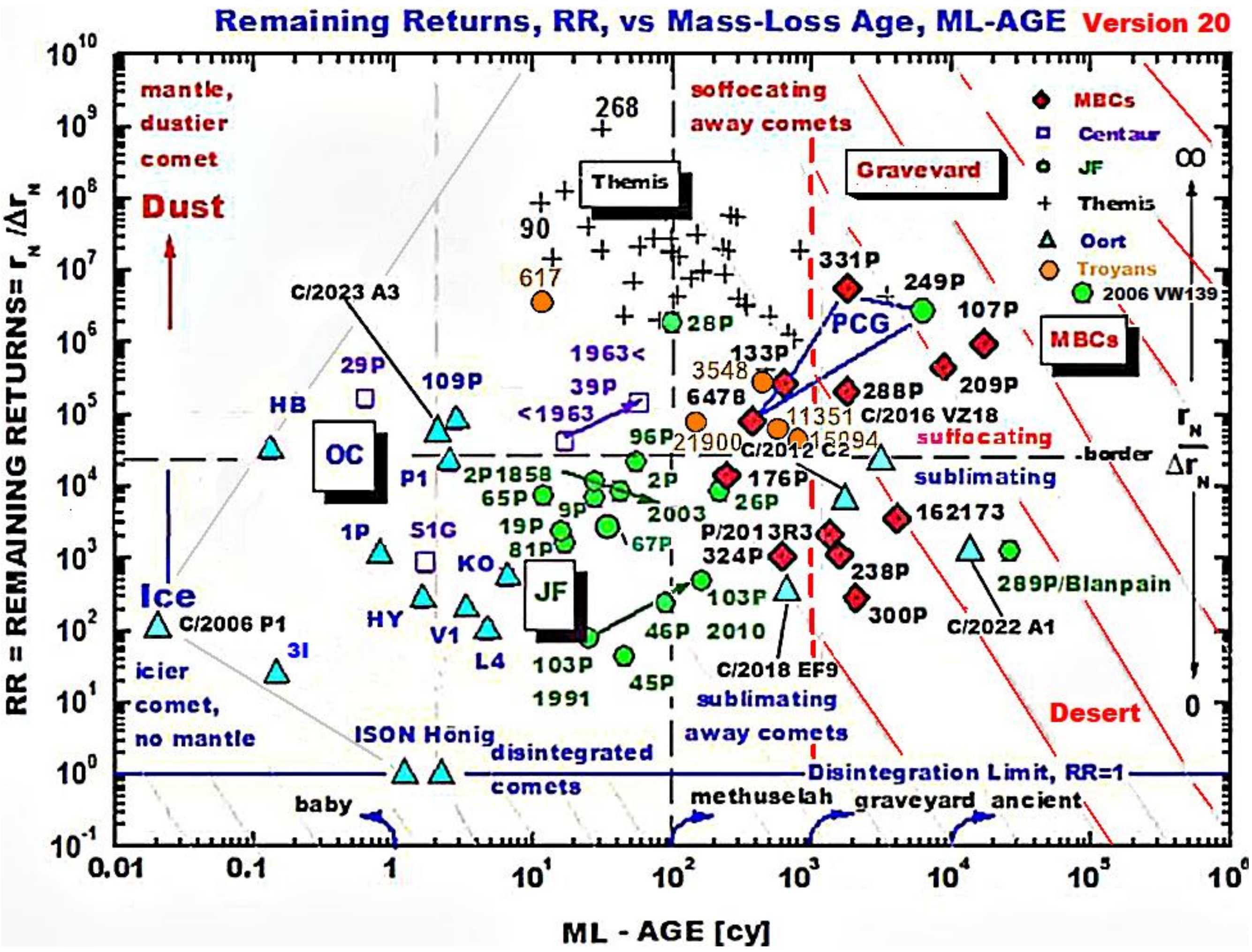

Figure 22. This is Version 20 of the Remaining Returns vs Mass-Loss Age diagram. The age is expressed in comet years which are different from Earth's years. 3I lies in the extreme lower left part of the diagram, occupied by icier bodies with no mantle, and surrounded by Oort Cloud comets. This agrees with the results found above, where we determined that the comet was 93% made of $CO_2$ ice and only 3.9% dust. For a detailed description see the text since this is a rather complex diagram.

The following can be said about the CED:

(1) The axis depicts Remaining Returns (RR) versus Mass-Loss Age. RR = r /Δr = 3 M/ΔM. The definition for Mass-Loss Age = constant/Total Mass Loss in comet years, different from Earth's years. The inverse of the total mass loss is a proxy for age.

(2) Because the diagram is log–log and covers 9 and 7 orders of magnitude along the *y*- and *x*-axis, the diagram is very forgiving. An error by a factor of 2 in any of the two variables changes the location of the data point by a very small amount.

(3) If a comet were made of pure ice, the layer removed by apparition $\Delta r$ would remain approximately constant (see Ferrín 2019, and Appendix 1), and $r/\Delta r \rightarrow 0$ as the comet sublimates away. If the comet contained much dust, part of it would fall back on the surface, $\Delta r$ would tend to zero, and $r/\Delta r \rightarrow \infty$. The same would happen if there were rocks embedded in the ice. As the surface recedes, the rocks cannot be lifted by the gas and would accumulate on the surface, once again choking further sublimation. Thus sublimating-away comets move down and to the right, and suffocating comets move up and to the right in the diagram.

(4) Necessarily there is a border that separates these two regimes, sublimating (below) and suffocating (above). The location of this border in the diagram was chosen just below pinpoint comet 6478 Gault, which belongs to a group of comets with very thin and long tails characteristic of suffocating comets, and above comet 2P/Encke that we know is sublimating in the right direction. The current value of this parameter is RR(Border) = 1E4 < 2.2E4 < 8.0E4.

(5) If a comet were to lie exactly on the border, it would move horizontally on the diagram. The following objects, Hale-Bopp, Halley, 96P and C/2012 C2 might be in this situation.

(6) The area below ML-AGE < 1 cy belongs to baby comets. The area above ML-AGE > 100 cy is for methuselah comets. The area with ML-AGE > 1000 cy is defined as the graveyard. And the region above ML-AGE > 10,000 cy is for ancient comets.

(7) Five comets in the graveyard are being suffocated: 331P, 249P, 107P, 209P and 288P.

(8) 2006 VW139 is the most extreme object in the upper right-hand corner with an age of 49000 comet years.

(9) Seven comets in the graveyard are still sublimating: C/2012 C2, 162173, P/2013 R3, 238P, 300P, C/2023 A1 and 289P. It is not surprising that ABCs occupy the upper right-hand corner of the diagram: this is expected on physical grounds, because they are old (large ML-AGE) and that they have a substantial dust or crust layer (large $r/\Delta r$).

(10) The CED shows that comets lie in well-defined groups. The Oort Cloud Comets to the left of the diagram (young and very young comets), the Jupiter Family comets (mature objects) lie at the center of the diagram, and the Main Belt Comets (old comets), lie to the right. The oldest comets lie completely on the right-hand side, the location of the graveyard.

(11) Comets scattered by Jupiter jump. Two comets had jumps in their orbits: Comet 39P and comet 103P. Comet 39P had a close encounter with Jupiter in 1963 and the orbit changed substantially. Fortunately, the SLC is known quite well, and it is possible not only to estimate the old mass loss but the future mass loss too. Comet 103P shows evolution between the 1991 and the 2010–2011 apparitions (Ferrín et al., 2012).

(12) Comet 2P/Encke is the only one that shows evolution of the parameters between the 1858 and the 2003 apparitions because there is an interval of 145 years. It was Kamel (1992) who created the 1858 light curve. All the other comets move too slowly or have shorter intervals and do no show motion.

(13) The diagram is a single frame of a movie that is playing at a very slow pace.

(14) If a comet decreases its perihelion distance, it receives more energy from the sun and it moves to the left and down rejuvenating, becoming a Lazarus Comets (Ferrín et al. 2013a).

(15) Since Lazarus comets move to the left and down on the diagram, and since a comet may enter the Lazarus-phase more than once in its lifetime, motion on the diagram will not be linear, and may exhibit backtracking and zig-zag paths. Not a single path has ever been calculated.

(16) If they approach the RR = 1 line (the last return), they sublimate completely or disintegrate, and comets ISON and Honing disintegrated.

(17) There should be a comet desert on the lower right-hand side of the diagram, where the objects would be very old and with few remaining returns. Since in this area comets sublimate with no residue left, their diameter is expected to be near zero and beyond detection, thus, we do not expect to find objects in this area, creating a desert. No objects are found, as expected, and if one appears it would require our immediate attention because it would be an extreme.

(18) RR = 1 is the disintegration limit. The comets plotted on the disintegration limit are C/2002 O4 Hönig, and C/2012 S1 ISON both of which in fact disintegrated (Sekanina, 2002). Many more could be plotted. In those cases in which the object disintegrated completely, it would be possible to calculate the size if the total mass loss can be ascertained and a density is chosen.

(19) Comet C/2006 P1 is the youngest comet in the diagram with Mass Loss Age = 0.02 comet years, a baby comet. Exocomet 3I is the second youngest equating comet Hale-Bopp but much more evolved, and Mass Loss Age = 0.16, another baby comet.

(20) Nomenclature: 1P, 1P/Halley; KO, C/1973 E1 Kohoutek; P1, C/2009 P1 Garradd; V1, C/2002 V1 NEAT; HB, C/1995 O1 Hale–Bopp; L4 = C/2011 L4 Panstars; HY, C/1996 B1 Hyakutake; C/2002 O4

Hönig; S1G, P/2009 S1 Gibbs. PCG=Pinpoint Comet Group.

Thus, the plot RR versus ML-AGE has the characteristics of an evolutionary diagram.

## 10- Conclusions

1- If we plot the location of 3I/ATLAS in the color-color diagrams, B-V vs V-R and V-R vs R-I, it lies among other comets of our solar system, thus it is a bona fide comet.

2- We produce several plots using SLC Methodology. The phase plot $m_V(1,1,\alpha)$ looks completely chaotic. No phase is discernible. This implies that we are not seeing the surface. The atmosphere is optically thick. The time plot allows the integration below the curve to get the total budget. And the log plot gives the value of several parameters cited below.

3- The SLC shows a photometric anomaly between -120 and -45 days, in the form of a diminishing magnitude that we interpreted as an eclipse. Thus, the object might be double.

4- From the log SLC we derive parameters ton= -371±10 d, while toff= +548±20 d, the dates of turn on and turn off, ton= 2024 10 14±10d and toff= 2027 05 01±20d. These correspond to solar distances Ron= -12.96±0.10 au and Roff=19.01±0.10 au, and temperatures Ton=89.7°±5°K and Toff=73.9°±5° K.

4- The composition is $CO_2$=93.23%, dust=3.9%, $H_2O$=2.66%, and CO=0.022%. These results are robust because they were derived from observations by many independent astronomers.

5- The SLC shows a slow increase in brightness at discovery, but at the level of -45 days before perihelion (R=2.5 au), the slope increases abruptly. We interpret this as the onset of water sublimation.

6- Integrating the total mass loss, gas and dust, we calculate the ML-AGE = 0.16±0.05 comet years, a baby comet. This result agrees with Ren et al. (2026) who concluded that 3I originated from the outer regions of its parent planetary disk (the Oort cloud of the stellar system).

7- Knowing the size and assuming a density, we can calculate the mass of the object and the mass lost. With that information it is also possible to calculate the number of Remaining Returns, RR=24. Then we can plot the 3I on the CED.

8- The CED can distinguish from what region of the stellar system the comet came from, the Jupiter family, the main belt, the Oort Cloud and the Pinpoint Comet Group, the Troyans and more. We find an interesting result, the comet belongs to the Oort Cloud, but of a different stellar system.

9- The CED is the first frame of a movie that is playing at a very slow pace.

10- 3I/ATLAS is depleted in dust and CO and enriched in $CO_2$. It has a $CO_2$ dominated coma, which may be explained partly by the preferential loss of CO over millennia because of its extreme volatility. The other possible explanation is that it was created in a dust-poor CO-poor location.

11- Exocomet 3I joins comets C/2006 R2, 29P and 103P as extreme objects compositionally.

12- The CED exhibits complexity beyond current understanding.

## 10- Comparison of Dimensions of CED vs HRD Phase Spaces

Since the Herzprung-Russell diagram is so famous, it is interesting to compare its size with the size of the CED diagram. This comparison is made in Figure 23. We find that the size of the CED diagram is about 8 times the size of the HR diagram.

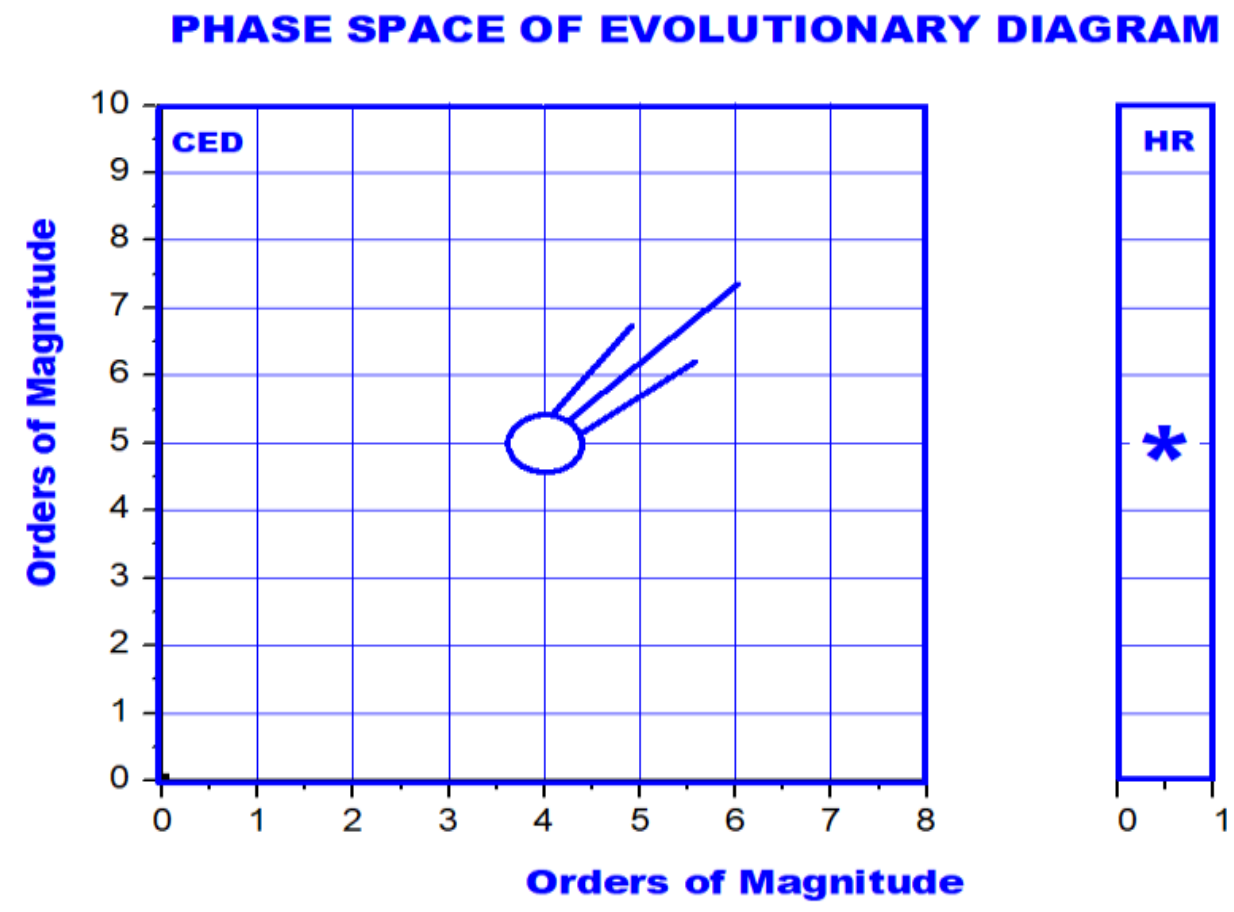

Figure 23. It is interesting to compare the sizes of the phase space of both diagrams, the CED and the HR-diagram. We see the CED is about 8 times larger than the HR.

## 11- Further research

The Comet Evolutionary Diagram, CED, needs additional research to grasp its full significance. Observationally, (1) more objects must be added, especially at the extremes, up (suffocated), down (disintegrated), left (young) and right (old), to know the limits of this phase space (see Section 10). Theoretically: (2) We need theoretical analysis to elucidate their trajectories in the plot, which we know will not be linear and may exhibit backtracking and zig-zag motion. The suffocation and sublimation regions will show differences. (3) Calculate the location of the border separating the suffocating comets from the sublimating comets. (4) Expand our understanding of the CED. For example, the CED displays the location of the graveyard but, is there a nursery? (5) We are seeing their current location after millions of years moving on the diagram. Where do they come from? What is needed is a forward and a backwards integration. (6) We need theoretical insight to understand why comets have a maximum production rate in Asec.

# Appendix 1- Layer Removed by Apparition, $\Delta r$, Approximately Constant vs Time

For a sublimating away comet, the thickness of the layer removed each apparition $\Delta \mathbf{r}$ should remain approximately constant as a function of time, as can be seen from the following argument. The energy captured from the Sun depends on the cross section of the nucleus, $\pi r^2$, on the Bond Albedo, AB, and on the solar constant, S. The energy conservation equation can be written:

$$(1- A_B)\, S\, \pi\, r_N^2 = p_{IR}.\sigma.T^4 + K1.4.\pi.\, r_N^2.\, \Delta r_N\, .L + K2\, \partial T/\partial x \quad (12)$$

where $A_B$ is the Bond albedo, S the solar constant, $r_N$ the nuclear radius, $p_{IR}$ is the albedo in the infrared, T the temperature, K1 and K2 are constants, and $\sigma$ the Stephan-Boltzmann constant, L the latent heat of sublimation, $\Delta r_N$ the thickness of the layer removed, x the depth below the surface. The term on the left is the energy captured from the Sun. The first term on the right is the energy radiated, the second the energy sublimated, and the third the energy conducted into the nucleus. The first and third terms on the right hand side are small in comparison with the second term because at large distances the temperature is very low. The second term dominates near the Sun at perihelion. So, in first approximation

$$(1- A_B)\, S\, \pi\, r_N^2 \sim K1\, .4.\, \pi.\, r_N^2.\, \Delta r_N\, .\, L \quad (13)$$

$$\Delta r_N \sim (1- A_B)\, S\, /\, 4\, K3\, L$$

We find that $\Delta r_N$ should be approximately constant, assuming that the orbit does not change, and the pole orientation remains stationary.

# ACKNOWLEDGMENTS

The authors acknowledge with thanks the COBS Comet Observations Database, contributed by observers worldwide and used in this research. The same acknowledgement is valid for the MPC observations database, with its hundreds of contributing observers. Their work allows the determination of the Secular Light Curves with great precision. In particular the work of the visual observers allows the extraction of the full flux from the coma, thus it must continue.

## 14- References